\documentclass[fleqn,usenatbib]{mnras}

\usepackage{newtxtext,newtxmath}
\usepackage[T1]{fontenc}

\usepackage{graphicx}
\usepackage{amsmath}
\usepackage{lastpage}

\DeclareRobustCommand{\VAN}[3]{#2}
\let\VANthebibliography\thebibliography
\def\thebibliography{%
  \DeclareRobustCommand{\VAN}[3]{##3}%
  \VANthebibliography
}

\title[Dust reverberation in Mrk~42 and Mrk~493]
{The 2MIG isolated AGNs -- 3. Optical--IR variability and dust reverberation
in the NLSy1 galaxies Mrk~42 and Mrk~493}

\author[I.O. Izviekova et al.]{
I.O. Izviekova,$^{1,2}$\thanks{E-mail: izviekova@gmail.com}
I.B. Vavilova,$^{1}$
O.V. Kompaniiets,$^{1}$
O. Zamora,$^{3,4}$
R. Clavero$^{3,4}$
\\
$^{1}$Main Astronomical Observatory of the NAS of Ukraine, 27, Akademik Zabolotny St., Kyiv, 03143, Ukraine\\
$^{2}$International Centre for Astronomical, Medical and Ecological Research (ICAMER), 27, Akademik Zabolotny St., Kyiv, 03143, Ukraine\\
$^{3}$Instituto de Astrofísica de Canarias, 38205 La Laguna, Spain\\ 
$^{4}$Departamento de Astrofísica, Universidad de la Laguna, 38206 La Laguna, Tenerife, Spain
}

\date{Accepted 2026 May 7. Received 2026 November 13; in original form 2026 November 13}

\pubyear{2026}

\begin{document}

\raggedbottom
\label{firstpage}
\pagerange{\pageref{firstpage}--\pageref{LastPage}}
\maketitle

\begin{abstract}
We present an optical--infrared variability study of two isolated narrow-line Seyfert 1 galaxies, Mrk 42 and Mrk 493, using ZTF optical and WISE/NEOWISE W1/W2 light curves, supplemented by ATLAS, SDSS, Swift, and IAC80 data. We analyse colour--magnitude and
flux--flux relations and measure optical--infrared reverberation lags
using flux-randomisation and random-subset-selection (FR/RSS)
realisations of the interpolated cross-correlation function
after correcting the WISE light curves for the variable accretion-disc
contribution. For Mrk~42, we obtain observed-frame lags of
$54^{+13}_{-33}$ and $62^{+7}_{-35}$~d in $W1$ and $W2$, respectively.
After applying the redshift/wavelength correction, these correspond
to dust-response radii of
$0.045^{+0.011}_{-0.028}$ and
$0.052^{+0.006}_{-0.029}$~pc.
For Mrk~493, the corresponding observed-frame lags are
$94^{+18}_{-21}$ and $111^{+26}_{-24}$~d, giving corrected
dust-response radii of
$0.078^{+0.015}_{-0.017}$ and
$0.092^{+0.021}_{-0.020}$~pc,
respectively. Both galaxies show optical bluer-when-brighter behaviour, whereas
their $W1-W2$ responses differ: the relation is significant in
Mrk~493 but not in Mrk~42, where host and circumnuclear dilution
may be stronger. For Mrk~42, SDSS spectral decomposition provides
a host-corrected active galactic nucleus (AGN) luminosity for the
contextual broad-line region (BLR) comparison.
The measured dust-response radii are broadly consistent with the
low-luminosity extension and intrinsic scatter of published
$R_{\rm dust}$--$L_{\rm bol}$ relations. The major optical flare in
Mrk~42 exhibits four broad internal maxima with
an intra-flare timescale of approximately 45--47~d;
non-uniform peak intervals and stochastic-model comparisons favour
flare substructure within a broader accretion-driven event.
Both galaxies follow the expected disc--hot-dust reverberation
framework while exhibiting distinct colour and optical-variability
phenomenology.
\end{abstract}

\begin{keywords}
accretion, accretion discs -- galaxies: active -- galaxies: Seyfert --
techniques: photometric -- infrared: galaxies -- galaxies: individual: Mrk~42, Mrk~493
\end{keywords}

\section{Introduction}
\label{sec:intro}

Multi-band variability provides a means of probing the unresolved
central regions of active galactic nuclei (AGNs), including the
accretion flow and its coupling to circumnuclear material
\citep{Peterson1993,Ulrich1997,Peterson2001}. The variable optical
and ultraviolet (UV) continuum is generally associated with the
accretion disc surrounding the supermassive black hole, while an
increasing fraction of the infrared emission arises from dust heated
by the primary continuum
\citep{Shakura1973,Peterson1997,Barvainis1987,Antonucci1993,
Netzer2015}.

Delays between the optical continuum and the infrared response
measure the characteristic scale of the hot-dust-emitting region.
Optical--infrared reverberation therefore connects the accretion
disc, the broad-line region (BLR), and the inner dusty structure
\citep{Blandford1982,Peterson1993,Peterson2004RM,
Suganuma2006,Koshida2014,Minezaki2019}.
The Wide-field Infrared Survey Explorer (\textit{WISE})
$W1$ and $W2$ bands, centred at 3.4 and
4.6~$\mu$m, are sensitive to hot dust near the inner boundary of
the circumnuclear dust distribution, and the observed flux can also
contain a variable accretion-disc contribution
\citep{Koshida2014,Mandal2021,Mandal2024DustTorus}, which is removed
in the lag analysis presented here.
At wavelengths beyond approximately
8--10~$\mu$m, spatially extended polar dust can contribute more
strongly \citep{Honig2013,Asmus2019}. The disc-corrected $W1/W2$
lags are therefore interpreted as effective response-weighted radii
of the hot inner dust.

WISE-based dust reverberation studies have extended such
measurements from nearby AGNs to large quasar samples.
\citet{Lyu2019DustRM} and \citet{Yang2020DustRM} established
well-defined infrared lag--luminosity relations over broad ranges
of luminosity and redshift, while \citet{Chen2023} and
\citet{Mandal2024DustTorus} quantified the wavelength-dependent
$W1/W2$ dust scales and their size--luminosity relations.
\citet{Ayubinia2025} further placed these mid-infrared (MIR) dust
scales in the context of the H$\beta$ BLR. These results provide the population
reference for our measurements of the two low-luminosity isolated
narrow-line Seyfert~1 galaxies.

The combination of high-cadence
Zwicky Transient Facility (ZTF) optical monitoring and
decade-long \textit{WISE} and Near-Earth Object
Wide-field Infrared Survey Explorer (\textit{NEOWISE})
photometry enables such measurements for nearby AGNs
\citep{Bellm2019ZTF,Masci2019ZTF,Graham2019,
Wright2010WISE,Mainzer2014NEOWISE,Cutri2012WISE,
Cutri2015NEOWISE}.
UV and optical continuum-reverberation studies show that
inter-band lags broadly follow the wavelength dependence expected
from disc reprocessing, although simple thin-disc size estimates
can underpredict the observed delays
\citep{Edelson2019,Guo2022,Cackett2020}.
Detailed thermal-reverberation modelling of X-ray-illuminated
Novikov--Thorne discs reproduces the observed UV--optical lag
spectra of monitored Seyfert galaxies
\citep{Kammoun2021}.
Updated calculations that explore a wider range of disc and
coronal parameters also reproduce the overall lag spectra, while
showing that individual bands can contain an additional diffuse
BLR contribution
\citep{Kammoun2023}.

Colour--magnitude and flux--flux diagnostics provide complementary
constraints on the variable continuum. Many AGNs show
bluer-when-brighter optical behaviour, consistent with a variable
component that is spectrally harder than the more slowly varying
central-region emission. Host-galaxy dilution and multiple variable
components can modify these relations
\citep{Giveon1999,Wilhite2005,Schmidt2012,Kokubo2014,Sun2014,
Choloniewski1981,Pereyra2006,Cai2016}. These effects are
particularly relevant for narrow-line Seyfert~1 galaxies (NLSy1s),
which commonly have relatively low black-hole masses, high
accretion rates, and strong optical Fe~II emission
\citep{Mathur2000,Zhou2006,Rakshit2017,Vasylenko2020,Waddell2022}.

The host-galaxy environment provides additional context for
interpreting nuclear variability within the broader evolutionary
state of a galaxy \citep{Vavilova2026MWNeTs}. Isolation reduces
the likelihood of strong ongoing tidal perturbations. AGNs selected
from the 2MASS Isolated Galaxies (2MIG) catalogue
\citep{Karachentseva2010} span a broad
range of nuclear and host-galaxy properties
\citep{Pulatova2015,Vavilova2016,Vasylenko2020,Pulatova2023,
Kompaniiets2023,Kompaniiets2025a,Kompaniiets2025b,
Dobrycheva2025}. Isolated systems therefore offer a useful setting
for examining disc--dust coupling in hosts with limited evidence
for recent strong external interactions.

We investigate the optical and infrared variability of the nearby
isolated NLSy1 galaxies Mrk~42 and Mrk~493, both commonly
classified as radio-quiet systems
\citep{Yang2020EESBHradio,Berton2020RQNLSy1,
SEAMBH2014RQNLSy1}. The primary analysis combines ZTF optical
photometry with WISE/NEOWISE $W1/W2$
monitoring. The Asteroid Terrestrial-impact Last Alert
System (ATLAS) extends the optical coverage of Mrk~493 and
provides a robustness test for its lag measurement. We examine
whether the nuclei show measurable optical--infrared delays
consistent with hot-dust reverberation and what their colour and
flux--flux relations reveal about the connection between the
variable optical continuum and the dust response.

Beyond measuring optical--infrared reverberation lags, we place the
inferred dust radii in the broader context of BLR scaling relations
and investigate whether additional time-domain diagnostics reveal
evidence for more complex variability processes.
For the radial-scale comparison, we use a common empirical
BLR radius--AGN luminosity at 5100~\AA\ relation
($R_{\rm BLR}$--$L_{5100}$)
\citep{WangWoo2024}.
For Mrk~42, the AGN continuum luminosity at
5100~\AA\ is derived through
Sloan Digital Sky Survey (SDSS) spectral decomposition. For
Mrk~493, the published direct H$\beta$ reverberation lag provides
an independent check on the relation-based BLR scale
\citep{Du2015SEAMBH4}. We also examine the internal structure of
the major optical flare in Mrk~42, while archival Swift
UV and X-ray observations provide complementary multi-band context
for Mrk~493.

Section~\ref{sec:sample_data} introduces the two targets and
summarises the optical, infrared, spectroscopic, UV, X-ray, and
dedicated ground-based data sets used in this study.
Section~\ref{sec:methods} describes the photometric preprocessing,
colour and flux--flux analyses, optical--infrared lag measurements,
spectral decomposition, and supporting time-domain methods.
Section~\ref{sec:results} presents the variability, reverberation,
and radial-scale results for Mrk~42 and Mrk~493.
Section~\ref{sec:discussion} interprets these results in the context
of accretion-disc variability, host-galaxy dilution, and hot-dust
reprocessing. The main conclusions are summarised in
Section~\ref{sec:conclusions}.

\section{Targets and data}
\label{sec:sample_data}

We analyse two nearby isolated narrow-line Seyfert~1 (NLSy1)
galaxies, Mrk~42 and Mrk~493, whose basic properties are listed in
Table~\ref{tab:objects}. Both sources lie at low redshift
($z<0.05$) and are classified as NLSy1 galaxies
\citep{Osterbrock1985,SEAMBH2014RQNLSy1,Pulatova2015,
Pulatova2023}. They are commonly treated as radio-quiet systems
\citep{Yang2020EESBHradio,Berton2020RQNLSy1,
SEAMBH2014RQNLSy1}. Their isolated hosts provide a useful setting
for examining AGN variability and disc--dust coupling in galaxies
with limited evidence for strong ongoing tidal perturbations
\citep{Vavilova2009,Karachentseva2010,Vavilova2021,
Kompaniiets2025b}. The adopted black-hole masses are taken from
\citet{Du2015SEAMBH4} for Mrk~493 and
\citet{Du2014BLRNLR} for Mrk~42. The host-galaxy inclination
angles are adopted from the HyperLEDA database
\citep{Paturel2003HyperLEDA}.

Mrk~42 is classified as an SB(r) galaxy
\citep{Vaucouleurs1991rc3,Nair2010}. Integral-field observations
reveal circumnuclear star-forming structure on sub-kpc scales
\citep{Hennig2018}. Mrk~493 is classified as an SB(r)b galaxy
\citep{Vaucouleurs1991rc3} and contains a bright nucleus surrounded
by star-forming structure detected in optical and ultraviolet imaging
and spectroscopy \citep{Munoz2007,Popovic2009}. These structures
lie within the central regions sampled by the unresolved survey
photometry and contribute to the measured unresolved emission. Physical scales and luminosity distances were calculated using the
Planck18 flat $\Lambda$CDM cosmology implemented in
\texttt{astropy}, with
$H_{0}=67.66~{\rm km~s^{-1}~Mpc^{-1}}$ and
$\Omega_{\rm m}=0.30966$
\citep{Planck2020}.

\begin{table*}
\centering
\caption{Basic properties of the isolated AGN host galaxies.}
\label{tab:objects}
\small
\begin{tabular}{@{}lccccccc@{}}
\hline
Object & RA & Dec & Morphology & Activity type &
$\log(M_\bullet/M_\odot)$ & $z$ & $i$ \\
       & (deg) & (deg) & & & & & (deg) \\
\hline
Mrk~493 (UGC~10120) &
239.790102 & 35.029861 & SB(r)b & NLSy1 &
6.14 &
$0.031503 \pm 1.03\times10^{-5}$ & 55.7 \\
Mrk~42 &
178.424042 & 46.211719 & SB(r) & NLSy1 &
6.37 &
$0.024251 \pm 3.51\times10^{-5}$ & 31.2 \\
\hline
\end{tabular}

\vspace{0.5ex}
\parbox{\textwidth}{\footnotesize
Note. Column $i$ gives the host-galaxy inclination. Morphological
types are taken from \citet{Vaucouleurs1991rc3} and
\citet{Nair2010}, where applicable. The black-hole masses are
adopted from \citet{Du2015SEAMBH4} for Mrk~493 and
\citet{Du2014BLRNLR} for Mrk~42. The inclination angles are
adopted from the HyperLEDA database
\citep{Paturel2003HyperLEDA}. The listed redshifts are spectroscopic values compiled by
the NASA/IPAC Extragalactic Database (NED)
\citep{NED}.
}
\end{table*}

\subsection{Primary photometric data: ZTF, ATLAS, and NEOWISE}
\label{subsec:survey_photometry}

The reverberation analysis uses public optical and infrared survey
photometry. ZTF provides $g$, $r$, and $i$ monitoring over
2018--2025
\citep{Bellm2019ZTF,Masci2019ZTF,Graham2019}. For Mrk~493,
ATLAS extends the optical baseline and provides a robustness test
for the lag measurement \citep{Tonry2018ATLAS}. The infrared data
comprise multi-epoch $W1$ and $W2$ photometry from
WISE/NEOWISE over 2014--2024
\citep{Wright2010WISE,Mainzer2014NEOWISE,Cutri2012WISE,
Cutri2015NEOWISE}.

The NEOWISE measurements within each observing window were
combined into epoch-averaged points. Successive windows are
separated by approximately six months, so shorter-timescale
infrared variability is unresolved.

Foreground Galactic reddening was obtained from the
Schlegel--Finkbeiner--Davis maps \citep{Schlegel1998}. The ZTF
$g$, $r$, and $i$ photometry was corrected using the
Pan-STARRS1 extinction coefficients tabulated by
\citet{Schlafly2011}. The band-specific $W1$ and $W2$ extinction
values were obtained from the NASA/IPAC Infrared Science Archive (IRSA) Dust Extinction Service using
the Schlafly--Finkbeiner prescription. All photometry was corrected
for Galactic extinction before the colour, spectral, flux--flux,
and lag analyses. The preprocessing and lag procedures are
described in Section~\ref{sec:methods}.

\subsection{Aperture scale and central-region interpretation}
\label{subsec:aperture_context}

The survey measurements do not spatially resolve the immediate
nuclear region. For the adopted cosmology, the angular scales are
approximately $0.51~{\rm kpc~arcsec^{-1}}$ for Mrk~42 and
$0.65~{\rm kpc~arcsec^{-1}}$ for Mrk~493. The ZTF seeing criterion, expressed as the full width at half
maximum (FWHM),
${\rm FWHM}\leq3\arcsec$, corresponds to projected scales of
up to approximately 1.5 and 2.0~kpc, respectively. For the WISE
profile-fit photometry, the catalogue profile-fit radius indicators
are $\texttt{w1fitr}=\texttt{w2fitr}=7.5\arcsec$. These values
provide a catalogue-based characteristic angular scale for the
profile-fit photometry of these unresolved sources. These angular
radii correspond to projected scales of approximately 3.8~kpc for
Mrk~42 and 4.9~kpc for Mrk~493.

These scales exceed the reported sizes of the circumnuclear
structures in both galaxies. The ZTF and WISE measurements therefore
sample the unresolved central region, including the AGN, inner
host-galaxy light, and circumnuclear stellar emission. The mean
optical and infrared fluxes, colours, and median spectral indices are
interpreted as observed central-region quantities. Correlated
variability, colour--magnitude slopes, flux--flux relations, and
optical--infrared delays are more sensitive to the variable nuclear
component superposed on the more slowly varying unresolved emission.
Residual host dilution can still affect the measured variability
amplitudes, colour relations, and correlation strength.

\subsection{Supporting data sets}
\label{subsec:supporting_data}

For Mrk~42, the SDSS $3\arcsec$ fibre spectrum is decomposed to
measure the AGN continuum luminosity at 5100~\AA, which is used in
the contextual comparison between the dust reverberation radius and
the BLR scale inferred from the empirical
$R_{\rm BLR}$--$L_{5100}$ relation.
For Mrk~493, we adopt the host-corrected continuum luminosity
$\log[\lambda L_{\lambda}(5100\,\text{\AA})_{\rm AGN}/
({\rm erg\,s^{-1}})]=43.11\pm0.08$
and the direct rest-frame H$\beta$ reverberation lag
$\tau_{\mathrm{H}\beta}=11.6^{+1.2}_{-2.6}$~d from
\citet{Du2015SEAMBH4}. The luminosity provides the same
relation-based BLR comparison used for Mrk~42, while the measured
H$\beta$ lag supplies an independent object-specific check.

Archival Swift observations provide supplementary UV and
X-ray coverage of Mrk~493. Mrk~42 was monitored with the
0.82-m IAC80 telescope and the CAMELOT2 imager in the
Johnson--Bessell $V$ band over three consecutive nights in May
2025, providing a test of optical intraday variability (IDV). The
IAC80 observing and calibration material is presented in
Appendix~\ref{app:iac80}. The SDSS spectral decomposition, the Swift data reduction,
and the IAC80 differential-photometry procedure are described in the
corresponding subsections of Section~\ref{sec:methods}.

\section{Methods}
\label{sec:methods}

The primary analysis comprises colour--magnitude and flux--flux
relations and optical--infrared reverberation lags. The SDSS
decomposition, optical inter-band lags, flare-substructure analysis,
archival Swift observations, and IAC80 monitoring provide
source-specific supporting constraints.
Appendix~\ref{app:lag_diagnostics} presents the lag-analysis
diagnostics, Appendix~\ref{app:lag_method_details} describes the
lag-analysis implementation, and Appendix~\ref{app:additional_methods}
provides additional methodological details.

\subsection{Survey photometry and preprocessing}
\label{subsec:survey_preprocessing}

\subsubsection{ZTF optical photometry}
\label{subsubsec:ztf_photometry}

We used ZTF $g$, $r$, and $i$ photometry obtained from
point-spread-function (PSF) fits
\citep{Bellm2019ZTF,Masci2019ZTF,Graham2019}. Because seeing
variations can change the relative nuclear and host-galaxy
contributions in spatially extended sources, we retained only
measurements with finite magnitudes and uncertainties satisfying
\[
{\rm FWHM}\leq3.0\arcsec,\qquad
\sigma_m\leq0.20~{\rm mag},\qquad
{\rm sep}\leq1.5\arcsec,
\]
where $\sigma_m$ is the magnitude uncertainty and ${\rm sep}$ is
the separation from the adopted nuclear position. The adopted
uncertainty threshold is conservative relative to the actual
photometric precision: after the seeing and positional cuts, the
largest single-exposure uncertainties are 0.023~mag for Mrk~42 and
0.019~mag for Mrk~493.

For Mrk~42, the major flare interval was excluded from the automatic
outlier rejection, and all measurements within this interval
passing the above quality cuts were retained. Robust median-absolute-deviation filtering was applied to
the quiescent baseline. The adopted filtering parameters and
additional seeing-sensitivity tests are given in
Appendix~\ref{app:survey_processing}.

The cleaned measurements were combined into nightly
inverse-variance-weighted mean magnitudes. Quasi-simultaneous
$g-r$ and $r-i$ pairs were constructed by matching nightly
measurements within $\Delta t\leq0.35$~d.

\subsubsection{ATLAS forced photometry}
\label{subsubsec:atlas_photometry}

We inspected the available ATLAS,
All-Sky Automated Survey for Supernovae (ASAS-SN),
Catalina Real-time Transient Survey (CRTS), and
Palomar Transient Factory (PTF)
photometry for both targets. For Mrk~493, the ATLAS $c$-band
light curve provided the only useful extension of the variable
optical driver. For Mrk~42, the optical--infrared response is
dominated by the isolated 2018--2019 flare, while the additional
archival measurements mainly extend the low-variability baseline.
An extended-driver test was therefore constructed for Mrk~493
alone.

The ATLAS measurements were cleaned using their formal flux
uncertainties and survey quality diagnostics and combined into
nightly inverse-variance-weighted mean flux densities. The
$c$-band nightly fluxes were transformed to the ZTF $g$-band
scale over the interval of temporal overlap using orthogonal distance regression (ODR) with uncertainties in both coordinates.

The transformed ATLAS measurements were merged with the ZTF
$g$-band light curve to form an extended optical driver. The
homogeneous ZTF light curve defines the fiducial analysis, while
the combined ATLAS+ZTF driver tests the sensitivity of the inferred
lags to the longer optical baseline.

\subsubsection{NEOWISE infrared photometry}
\label{subsubsec:neowise_photometry}

We used NEOWISE single-exposure profile-fit photometry in the
$W1$ and $W2$ bands from the IRSA time-series products
\citep{Wright2010WISE,Mainzer2014NEOWISE,Cutri2012WISE,
Cutri2015NEOWISE}. The catalogue magnitudes are given in the Vega
system. The WISE PSF has an FWHM of approximately $6\arcsec$, so
the measurements are interpreted within the central-region
framework of Section~\ref{subsec:aperture_context}.

Measurements were retrieved within $5\arcsec$ of the adopted
nuclear coordinates and filtered using the NEOWISE signal-to-noise,
photometric-quality, artefact, and saturation diagnostics.
Consecutive measurements separated by no more than 10~d were
grouped into observing epochs, within which additional robust
outlier rejection was applied separately to $W1$ and $W2$.
The complete quality cuts and clipping criteria are given in
Appendix~\ref{app:survey_processing}.

Inverse-variance-weighted mean magnitudes were calculated for each
epoch. Successive NEOWISE epochs are typically separated by
approximately six months, so the resulting light curves trace
long-term infrared variability.

\begin{figure*}
\centering
\includegraphics[width=\textwidth]
{\detokenize{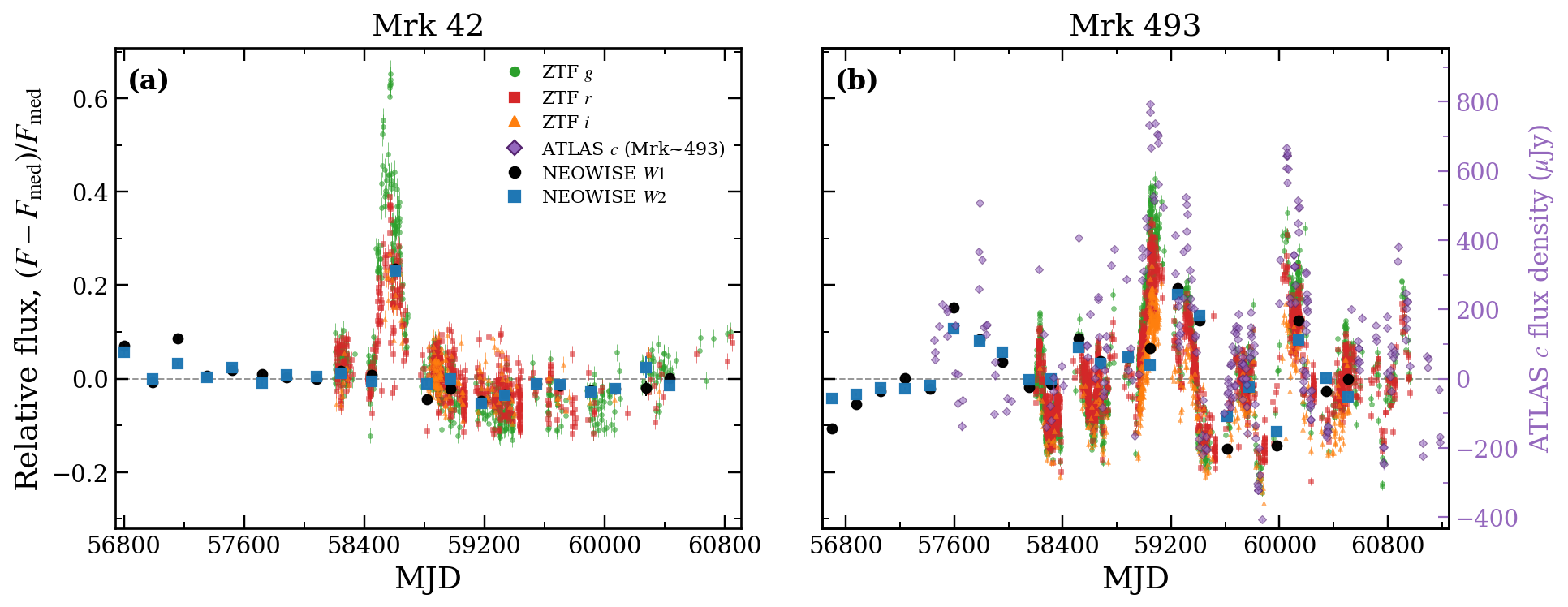}}

\caption{
Long-term optical and infrared variability of Mrk~42 and
Mrk~493. Panels (a) and (b) show the ZTF $g$, $r$, and $i$
measurements together with the epoch-averaged NEOWISE $W1$ and
$W2$ measurements for Mrk~42 and Mrk~493, respectively.
The ZTF and NEOWISE light curves are expressed as relative flux,
$(F-F_{\rm med})/F_{\rm med}$, using the median level of each
band. Panel (b) additionally shows the native ATLAS $c$-band
forced-photometry measurements, with their flux-density scale
given by the right-hand axis. All measurements are shown at their
observed epochs; no temporal shifts have been applied.
}
\label{fig:longterm_lc}
\end{figure*}

When AB colours or spectral indices were required, we applied
$W1_{\rm AB}=W1_{\rm Vega}+2.699$ and
$W2_{\rm AB}=W2_{\rm Vega}+3.339$
\citep{Cutri2012WISE}. The $W1-W2$ versus $W1$ regressions were
performed in the native Vega system; the additive transformations
do not affect the fitted slopes.
The long-term optical and epoch-averaged NEOWISE light curves
of both objects are shown in Fig.~\ref{fig:longterm_lc}, together
with the native ATLAS $c$-band measurements for Mrk~493.

\subsection{Colour and flux--flux diagnostics}
\label{subsec:colour_flux_methods}

\subsubsection{Colour--magnitude relations and spectral indices}
\label{subsubsec:colour_spectral_indices}

Optical colour--magnitude diagrams were constructed from the
quasi-simultaneous nightly ZTF pairs, using $g-r$ versus $g$ and
$r-i$ versus $r$. The infrared relation was constructed from the
epoch-averaged $W1-W2$ and $W1$ measurements. These colours
describe the unresolved central-region emission defined in
Section~\ref{subsec:aperture_context}.

Each relation was fitted with York regression
\citep{York2004}, which explicitly accounts for uncertainties in
both coordinates and their covariance. Because the magnitude plotted
on the abscissa also enters the corresponding colour, the covariance
between the two coordinates was included in the fit. We also
calculated the Pearson correlation coefficient and its associated
$p$-value.

We adopt the convention $F_{\nu}\propto\nu^{-\alpha}$. For two
filters $X$ and $Y$ in the AB system,
\begin{equation}
\alpha_{XY}
=
\frac{X-Y}{K_{XY}},
\qquad
K_{XY}
=
2.5\log_{10}
\left(
\frac{\lambda_Y}{\lambda_X}
\right),
\label{eq:alpha_colour}
\end{equation}
where $\lambda_X$ and $\lambda_Y$ are the effective wavelengths.
For a fitted colour--magnitude slope
$m_{\rm colour}=d(X-Y)/dX$,
\begin{equation}
\frac{d\alpha_{XY}}{d(-X)}
=
-\frac{m_{\rm colour}}{K_{XY}}.
\label{eq:alpha_slope}
\end{equation}

\subsubsection{Flux--flux relations}
\label{subsubsec:flux_flux_methods}

Flux--flux analysis was used to characterise the effective spectrum
of the dominant variable component
\citep{Choloniewski1981,PozoNunez2012,PozoNunez2013,
Ramolla2015,Minezaki2019}. A slowly varying background contributes
primarily to the intercept of an approximately linear relation,
while the slope describes the colour of the dominant correlated
variable component.

The extinction-corrected ZTF magnitudes were converted to flux
densities using
$F_{\nu}=3631~{\rm Jy}\,10^{-0.4m_{\rm AB}}$
\citep{OkeGunn1983}.
For each optical pair, we fitted
\begin{equation}
F_{\nu}(X)
=
a_{XY}\,F_{\nu}(Y)+b_{XY},
\label{eq:flux_flux}
\end{equation}
with $(X,Y)=(g,r)$ and $(r,i)$. Orthogonal distance regression was
used to account for uncertainties in both flux densities. For a correlated variable component described by
$F_{\nu}\propto\nu^{-\alpha_{\rm var}}$, the fitted flux--flux
slope was converted to a spectral index as
$\alpha_{\rm var}=-\ln a_{XY}/\ln(\nu_X/\nu_Y)$.

The same procedure was applied to the epoch-averaged WISE
measurements with $(X,Y)=(W1,W2)$. Vega magnitudes were converted
using $F_{\nu}=F_{\nu,0}10^{-0.4m_{\rm Vega}}$, with
$F_{\nu,0}=309.54$~Jy for $W1$ and
$F_{\nu,0}=171.787$~Jy for $W2$
\citep{Cutri2012WISE}. The infrared flux--flux relation therefore
characterises the longer-timescale variable component sampled by
the epoch-averaged NEOWISE observations.

The colour and flux--flux analyses use the observed $W1$ and $W2$
light curves. The light curves corrected for the variable
accretion-disc contribution are used exclusively in the
optical--infrared lag analysis.

\subsection{Optical--infrared lag analysis}
\label{subsec:lag_methods}

The optical--infrared delays were measured by cross-correlating the
cleaned ZTF $g$-band light curve with the epoch-averaged NEOWISE
$W1$ and $W2$ light curves. We primarily used the interpolated
cross-correlation function (ICCF)
\citep{Gaskell1987,White1994,Peterson1998,Peterson2004RM}.
As an independent consistency check, we also applied the
$z$-transformed discrete correlation function using pyZDCF
\citep{Alexander1997}.

\subsubsection{Accretion-disc correction}
\label{subsubsec:ad_correction}

The observed $W1$ and $W2$ fluxes can contain a variable
accretion-disc contribution. We estimated this component using the
spectral prescription applied by \citet{Mandal2024DustTorus}. The
ZTF $g$-band flux density and its uncertainty were linearly
interpolated to each NEOWISE epoch within the temporal coverage of
the optical light curve. No extrapolation beyond the optical
coverage was used.

For an infrared band with effective frequency $\nu_{\rm IR}$, the
estimated accretion-disc contribution is
\begin{equation}
F_{\nu,{\rm AD}}(\nu_{\rm IR},t)
=
F_{\nu,g}(t)
\left(
\frac{\nu_{\rm IR}}{\nu_g}
\right)^{s_{\rm AD}},
\label{eq:ad_component}
\end{equation}
where $\nu_g$ is the effective frequency of the ZTF $g$ band. The
corrected dust light curve is
\begin{equation}
F_{\nu,{\rm dust}}(\nu_{\rm IR},t)
=
F_{\nu,{\rm obs}}(\nu_{\rm IR},t)
-
F_{\nu,{\rm AD}}(\nu_{\rm IR},t).
\label{eq:dust_only_flux}
\end{equation}
The uncertainty of the corrected flux includes the WISE
measurement uncertainty and the propagated uncertainty of the
interpolated optical contribution.
The fiducial correction adopts $s_{\rm AD}=1/3$, corresponding to
the long-wavelength spectrum of a standard optically thick
accretion disc \citep{Koshida2014}. The cases $s_{\rm AD}=0.1$
and $s_{\rm AD}=0$ were evaluated as sensitivity tests for both
targets \citep{Minezaki2019}.

\subsubsection{Lag measurements and consistency checks}
\label{subsubsec:iccf_frrss}

The ICCF was calculated with bidirectional interpolation: each
light curve was interpolated in turn, and the two correlation
coefficients were averaged.
The lag grid is defined in the observer frame.
The lag grid covered
$-200\leq\tau\leq250$~d with a spacing of 1~d. The primary
positive-lag solution was identified within
$0\leq\tau\leq220$~d, and each ICCF point was required to contain
at least eight overlapping pairs.

For every band pair, we measured the peak lag, $\tau_{\rm peak}$,
and the centroid lag, $\tau_{\rm cent}$. The centroid was calculated
over the contiguous region around the primary peak satisfying
$r(\tau)\geq0.8\,r_{\rm max}$.
We adopted $\tau_{\rm cent}$ as the fiducial lag, while
$\tau_{\rm peak}$ was retained only as a diagnostic quantity.

Uncertainties in the fiducial optical--infrared lag measurements were
determined with 5000 flux-randomisation and random-subset-selection
(FR/RSS) realisations
\citep{Peterson1998,Peterson2004RM}. In each realisation, the fluxes
were perturbed according to their measurement uncertainties and
the light curves were resampled. The 16th, 50th, and 84th
percentiles of the centroid distribution define the reported lag
and its uncertainty. Convergence was checked explicitly after
2000, 4000, and 5000 realisations.

For both targets and both WISE bands, the FR/RSS centroid
distributions obtained for $s_{\rm AD}=0.1$ and $s_{\rm AD}=0$
overlapped the corresponding fiducial $s_{\rm AD}=1/3$ intervals.
We therefore adopt $s_{\rm AD}=1/3$ for the reported lag
measurements.

For Mrk~42, the 5000-realisation ICCF/FR/RSS diagnostics are shown in
Fig.~\ref{fig:app_mrk42_mirlags}. The $W1$ and $W2$ centroid
distributions remain stable between 2000, 4000, and 5000 FR/RSS
realisations (Fig.~\ref{fig:mrk42_frrss_convergence}).
The centroid-based ICCF/FR/RSS analysis was also repeated after
subtracting the fitted 47.12-d residual component from the optical
driver over Modified Julian Date (MJD) 58423.0--58692.0
(see Sections~\ref{subsec:flare_methods} and
\ref{sec:mrk42_flare_qpo} for the flare-substructure analysis
and results).
Paired FR/RSS realisations were used to quantify the corresponding
centroid shifts. The paired-realisation procedure is described in
Appendix~\ref{app:lag_method_details}.
Given the approximately six-month spacing of the NEOWISE epochs,
the pyZDCF results are used as qualitative consistency checks on the
ICCF solutions. The pyZDCF settings are given in
Appendix~\ref{app:lag_method_details}, and the corresponding
diagnostic curves are shown in
Fig.~\ref{fig:app_pyzdcf_all} of
Appendix~\ref{app:lag_diagnostics}.

For Mrk~493, the ICCF and FR/RSS analysis was repeated with 5000
realisations for both the fiducial ZTF-only driver and the combined
ATLAS+ZTF optical driver described in
Section~\ref{subsubsec:atlas_photometry}. The centroid distributions
were additionally compared after 2000, 4000, and 5000 realisations
to verify convergence. The inferred $W1$ and $W2$ lags remain stable
with increasing numbers of realisations and with extension of the
optical baseline. The corresponding distributions are shown in
Fig.~\ref{fig:app_mrk493_iccf}.

Following \citet{Mandal2024DustTorus}, we account for both
cosmological time dilation and the wavelength dependence of the
dust-reverberation radius. For
$R_{\rm dust}\propto\lambda^{\gamma}$ with $\gamma=0.62$, the
corrected lag is
\begin{equation}
\tau_{\rm corr}
=
\tau_{\rm obs}(1+z)^{\gamma-1}
=
\tau_{\rm obs}(1+z)^{-0.38}.
\label{eq:tau_corr}
\end{equation}
The corresponding response-weighted dust radius is
\begin{equation}
R_{\rm dust}
=
c\,\tau_{\rm corr}.
\label{eq:rdust}
\end{equation}

\subsection{SDSS decomposition and AGN continuum luminosity of Mrk~42}
\label{subsec:l5100_methods}

The SDSS $3\arcsec$ fibre spectrum of Mrk~42 was decomposed to
measure the AGN continuum luminosity at 5100~\AA\
\citep{York2000SDSS}. The fibre radius of $1.5\arcsec$ corresponds
to approximately 0.76~kpc and includes the AGN continuum, inner
host-galaxy starlight, and circumnuclear stellar emission.
The spectrum was fitted with
penalized pixel-fitting (pPXF) and
extended Medium-resolution Isaac Newton Telescope Library
of Empirical Spectra (E-MILES) stellar-population
templates \citep{Cappellari2017,Vazdekis2016}. The model comprised
a stellar component, a power-law AGN continuum, and an Fe~II
pseudo-continuum
\citep{VeronCetty2004,BorosonGreen1992}. The fit was performed
over the wavelength interval corresponding to 3750--6800~\AA\ in
the rest frame, with the strongest emission-line regions masked.

The fitted power-law flux density corresponding to rest-frame
5100~\AA\ was sampled at
$\lambda_{\rm obs}=5100(1+z)$~\AA\ in the original
observed-frame flux-density scale. The AGN continuum luminosity is
\begin{equation}
\lambda L_{\lambda}(5100\,\text{\AA})_{\rm AGN}
=
4\pi D_{\rm L}^{2}\,
\lambda_{\rm obs}
F_{\lambda,{\rm AGN}}^{\rm obs}(\lambda_{\rm obs}).
\label{eq:l5100_agn}
\end{equation}

The decomposition was repeated with an expanded emission-line mask.
The difference between the resulting luminosities was adopted as the
systematic uncertainty, giving
$\log[\lambda L_{\lambda}(5100\,\mbox{\AA})_{\rm AGN}/
({\rm erg\,s^{-1}})] = 42.55\pm0.07$.

\subsection{Supporting time-domain analyses}
\label{subsec:supporting_methods}

\subsubsection{Optical inter-band lags in Mrk~42}
\label{subsec:interband_methods}

The cleaned ZTF $g$, $r$, and $i$ light curves of Mrk~42 were
analysed at the level of individual exposures for
state-dependent optical continuum delays using the same
centroid-based ICCF/FR/RSS framework. These measurements are used
to interpret the state-dependent $r-i$ colour relation. The adopted lag ranges, sampling criteria, FR/RSS settings, and
pyZDCF checks are given in
Appendix~\ref{app:lag_method_details}; the corresponding flare-state
lag diagnostics are shown in
Fig.~\ref{fig:app_mrk42_optlags} of
Appendix~\ref{app:lag_diagnostics}.

\subsubsection{Mrk~42 flare substructure}
\label{subsec:flare_methods}

The internal structure of the major Mrk~42 optical flare was
examined over MJD 58423.0--58692.0.
After detrending the broad flare evolution, the residual structure
was characterised using generalised Lomb--Scargle (GLS) analysis
\citep{Lomb1976,Scargle1982,Zechmeister2009,VanderPlas2018},
peak-spacing measurements, Timmer--Koenig red-noise simulations
\citep{Timmer1995,Vaughan2005RedNoise}, and time-domain
Gaussian-process (GP) models implemented with \texttt{celerite2}
\citep{ForemanMackey2017Celerite}. The Gaussian-process comparison
included a stochastic exponential covariance model and a model
with an additional damped simple-harmonic-oscillator (SHO)
component; their relative support was assessed using the Akaike
information criterion (AIC) and Bayesian information criterion
(BIC) \citep{Liddle2007}. These tests were used to distinguish a
characteristic intra-flare timescale from a coherent periodic
interpretation. The detailed detrending, periodogram,
surrogate-light-curve, peak-timing, and Gaussian-process procedures
are described in Appendix~\ref{app:flare_method_details}.

\subsubsection{Swift observations of Mrk~493}
\label{subsec:swift_methods}

Archival Swift Ultraviolet/Optical Telescope (UVOT) and
X-Ray Telescope (XRT) observations were used to
provide UV/X-ray context for the long-term activity of Mrk~493.
The XRT analysis used the UK Swift Science Data Centre (UKSSDC) 0.3--10~keV products, while UVOT
photometry was extracted with \texttt{uvotsource}
\citep{Evans2007,Evans2009,Poole2008,Breeveld2011}. For the multi-band comparison, each series was expressed as a fractional deviation from its own median. The detailed screening, temporal grouping, aperture
photometry, and normalisation procedures are given in
Appendix~\ref{app:swift_processing}.

\subsubsection{IAC80 monitoring of Mrk~42}
\label{subsec:iac80_methods}

The IAC80/CAMELOT2 $V$-band monitoring of Mrk~42 was analysed with
differential aperture photometry using four local comparison stars.
Intraday variability was assessed with the power-enhanced $F$-test
at $\alpha=0.01$ \citep{deDiego2014EnhancedF}. The complete
photometric procedure, aperture definition, comparison-star
selection, observing log, and calibration material are presented in
Appendix~\ref{app:iac80}.
The observing log and differential light curves are given in
Table~\ref{tab:iac80_log} and Fig.~\ref{fig:mrk42_idv},
respectively. The comparison-star sequence and finding chart are
available in the \textit{AGN Reference Fields} dataset
\citep{Izviekova2026AGNfields}.

\section{Results}
\label{sec:results}

The results are organised around the diagnostics shared by
Mrk~42 and Mrk~493: long-term optical and WISE $W1/W2$
variability, colour and flux--flux behaviour, optical--infrared
reverberation lags, and the resulting hot-dust scales. The
source-specific flare, intraday, UV, and X-ray analyses are
presented subsequently as complementary time-domain constraints.

\subsection{Long-term optical and W1/W2 variability}
\label{subsec:results_longterm}

Both nuclei show significant long-term variability in the optical
and WISE bands. The epoch-averaged $W1$ and $W2$ light curves vary
more smoothly than the optical light curves, as expected for a
temporally broadened response from circumnuclear dust. Mrk~42 exhibits a prominent structured optical flare, whereas
Mrk~493 shows a more gradual long-term optical evolution
accompanied by a delayed infrared response.
The long-term ZTF, ATLAS, and WISE/NEOWISE light curves are shown
in Fig.~\ref{fig:longterm_lc}.

The common phenomenology motivates the optical--infrared
reverberation analysis below. The colour and flux--flux diagnostics
first establish how the unresolved central-region spectra evolve
with brightness and whether the correlated variable components
retain approximately stable spectral shapes.

\subsection{Colour--magnitude and flux--flux behaviour}
\label{subsec:results_colours_flux}

\subsubsection{Mrk~42}
\label{sec:mrk42_colours}

\begin{figure*}
\centering

\begin{minipage}{0.32\textwidth}
\centering
\includegraphics[width=\linewidth]
{\detokenize{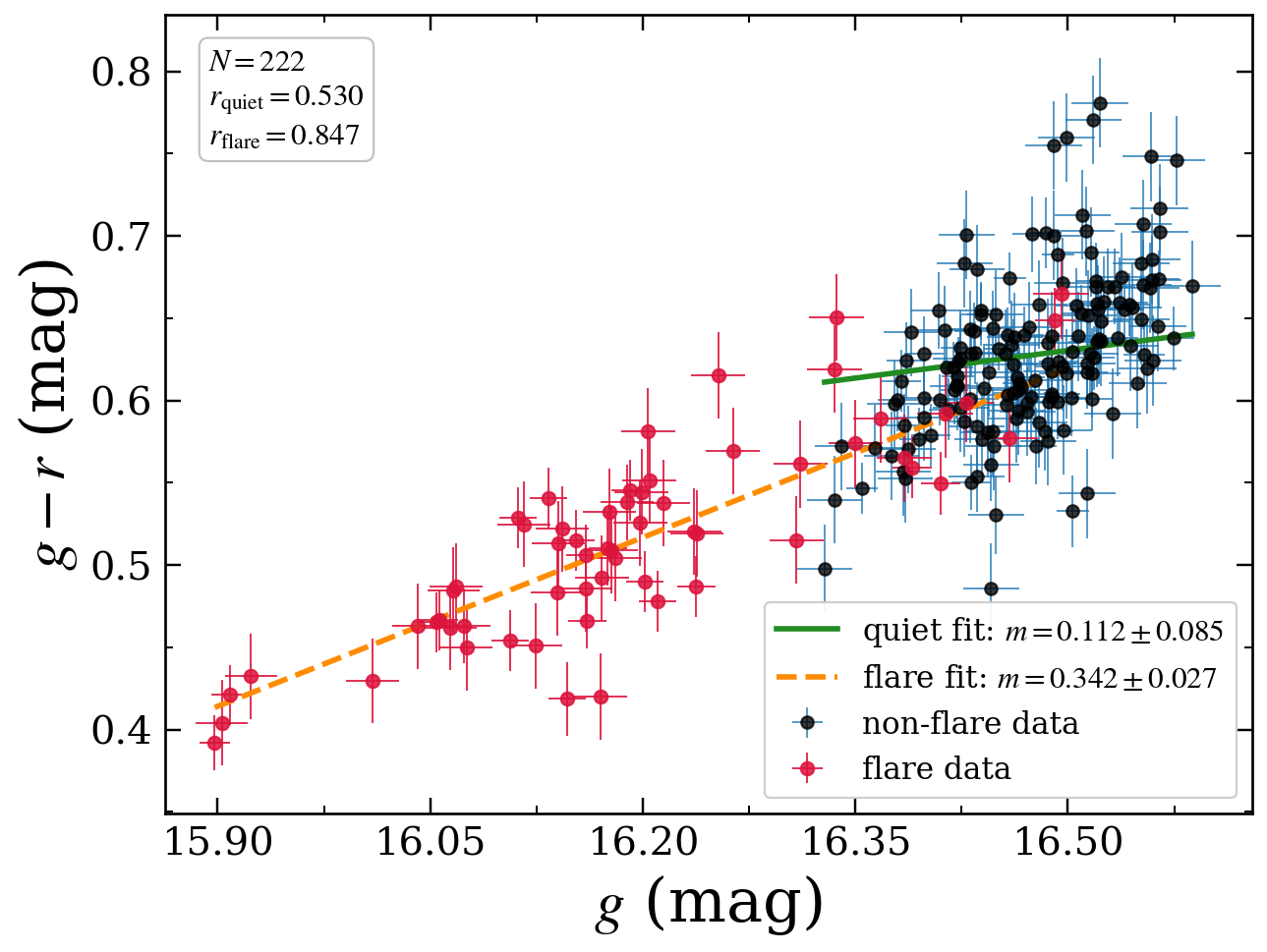}}\\[-0.5ex]
(a)
\end{minipage}
\hfill
\begin{minipage}{0.32\textwidth}
\centering
\includegraphics[width=\linewidth]
{\detokenize{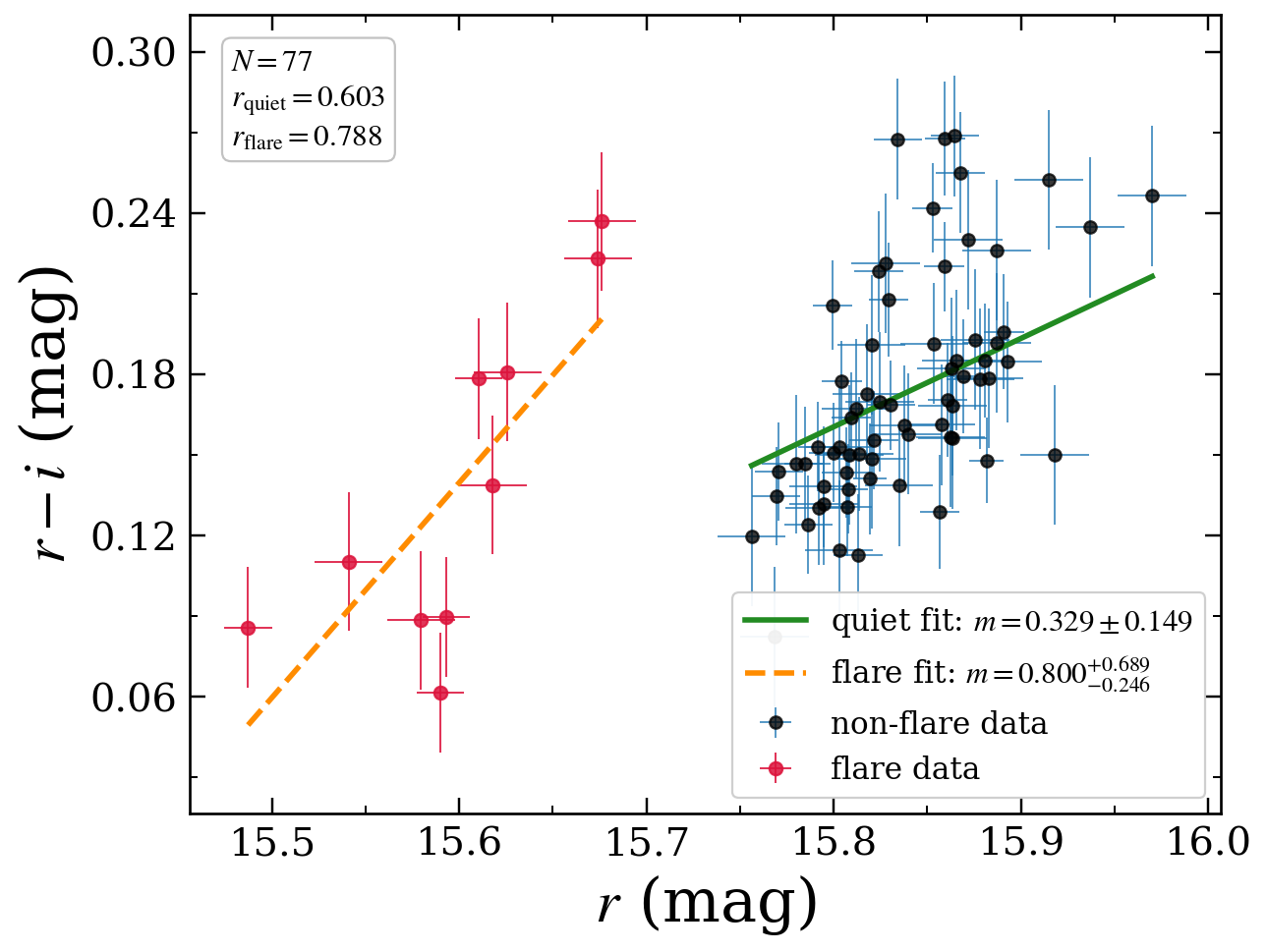}}\\[-0.5ex]
(b)
\end{minipage}
\hfill
\begin{minipage}{0.32\textwidth}
\centering
\includegraphics[width=\linewidth]
{\detokenize{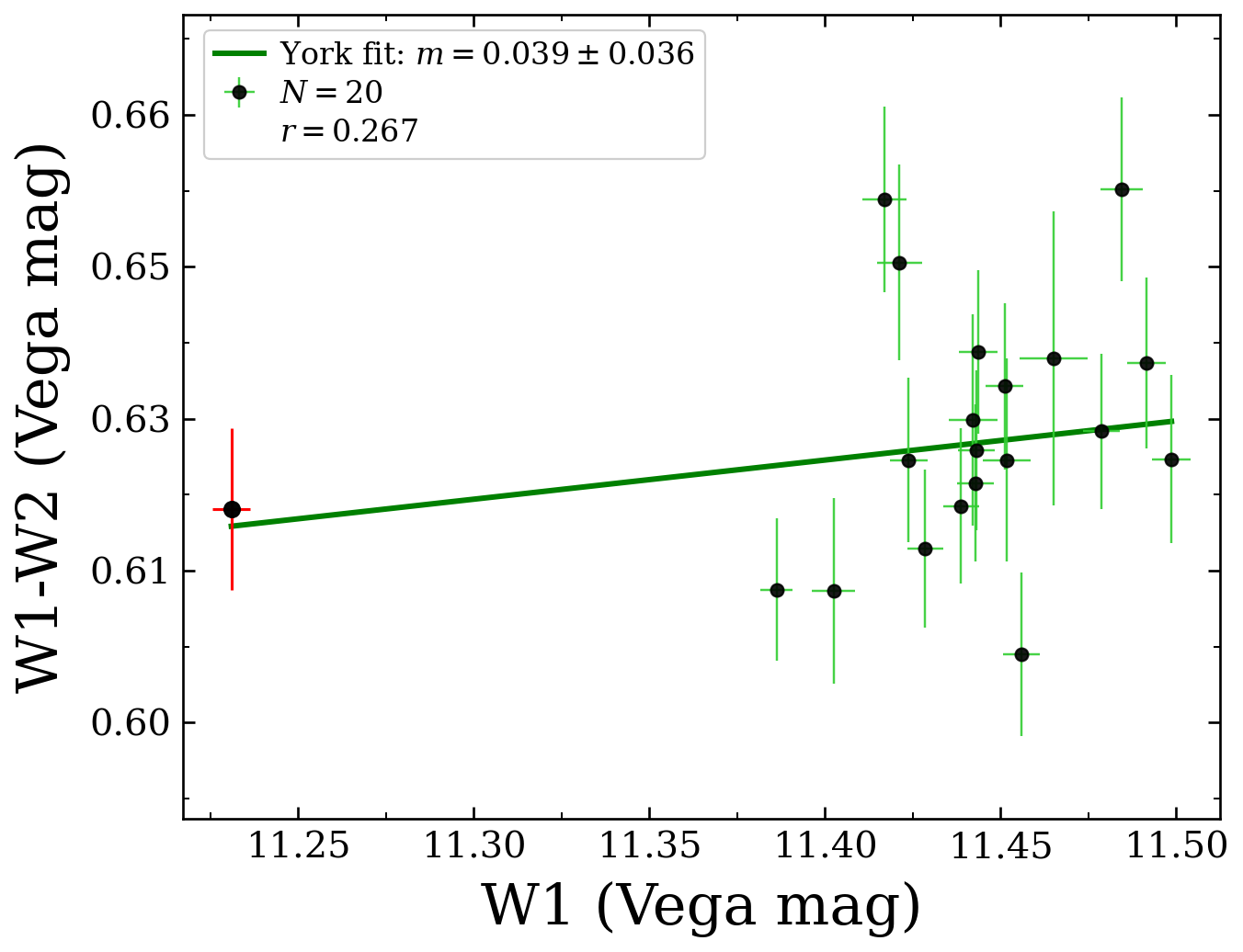}}\\[-0.5ex]
(c)
\end{minipage}

\caption{Colour--magnitude relations for Mrk~42.
Panel (a) shows $g-r$ versus $g$, panel (b) shows
$r-i$ versus $r$, and panel (c) shows $W1-W2$ versus $W1$.
The optical panels distinguish the quiet and flare intervals.
The measured colours describe unresolved central-region emission
and include contributions from the variable nucleus, host galaxy,
and circumnuclear emission.}
\label{fig:mrk42_colors}
\end{figure*}

The optical colour--magnitude relations of Mrk~42 show clear
bluer-when-brighter behaviour
(Fig.~\ref{fig:mrk42_colors}).
The quiet and flare intervals occupy distinct brightness
ranges and were analysed separately. Both the $g-r$ versus
$g$ and $r-i$ versus $r$ relations have positive slopes in
each state. The $g-r$ slope increases from
$m=0.112\pm0.085$ in the quiet state to
$m=0.342\pm0.027$ during the flare. The $r-i$ relation
is also steeper during the flare, although its slope has a
larger uncertainty owing to the smaller number of paired
measurements.
The epoch-averaged $W1-W2$ versus $W1$ relation shows no
statistically significant colour trend.
The fitted slopes, Pearson coefficients, median colours,
and colour-derived spectral indices are listed in
Table~\ref{tab:mrk42_spec}.

\begin{figure*}
\centering

\begin{minipage}{0.32\textwidth}
\centering
\includegraphics[width=\linewidth]
{\detokenize{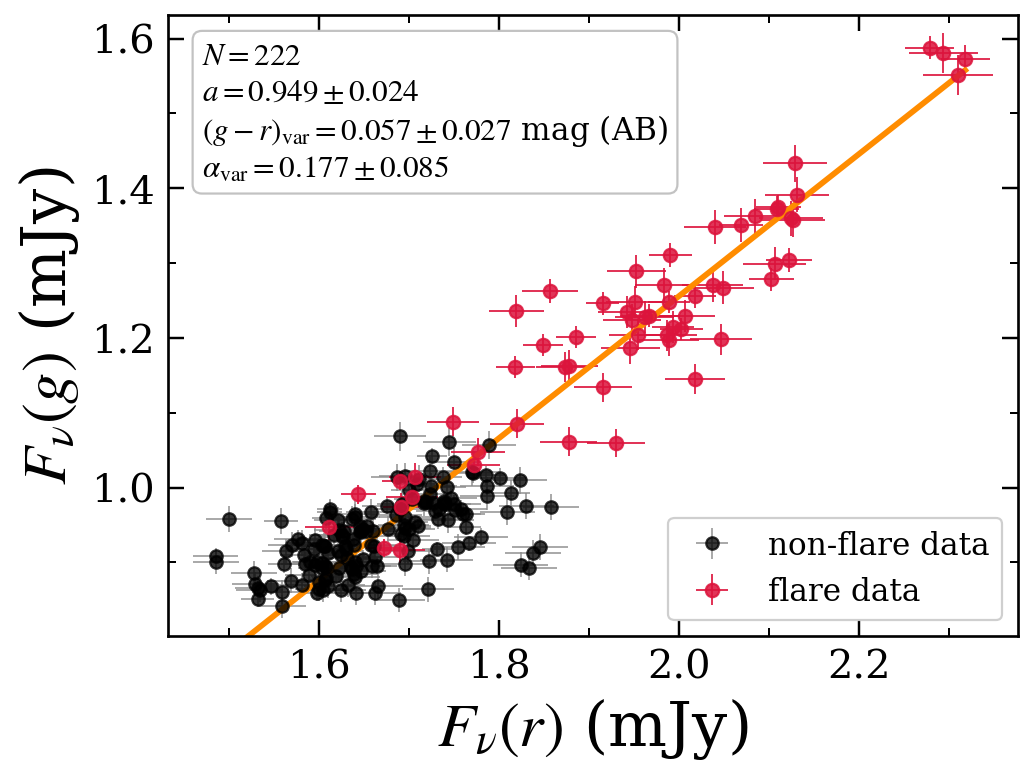}}\\[-0.5ex]
(a)
\end{minipage}
\hfill
\begin{minipage}{0.32\textwidth}
\centering
\includegraphics[width=\linewidth]
{\detokenize{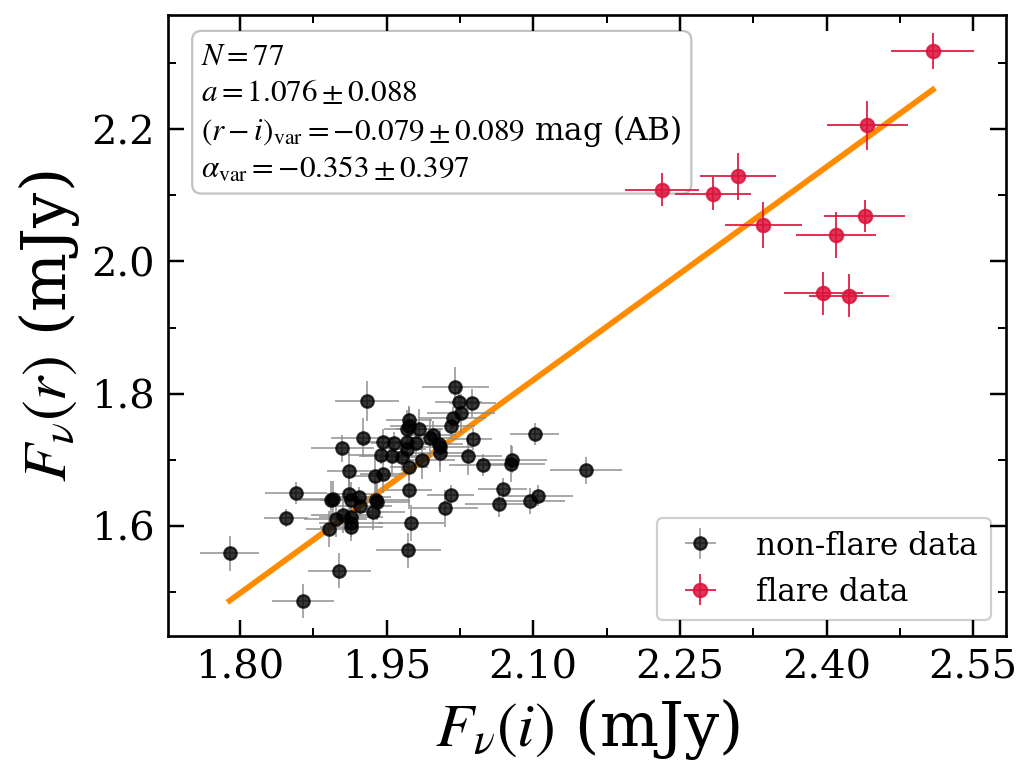}}\\[-0.5ex]
(b)
\end{minipage}
\hfill
\begin{minipage}{0.32\textwidth}
\centering
\includegraphics[width=\linewidth]
{\detokenize{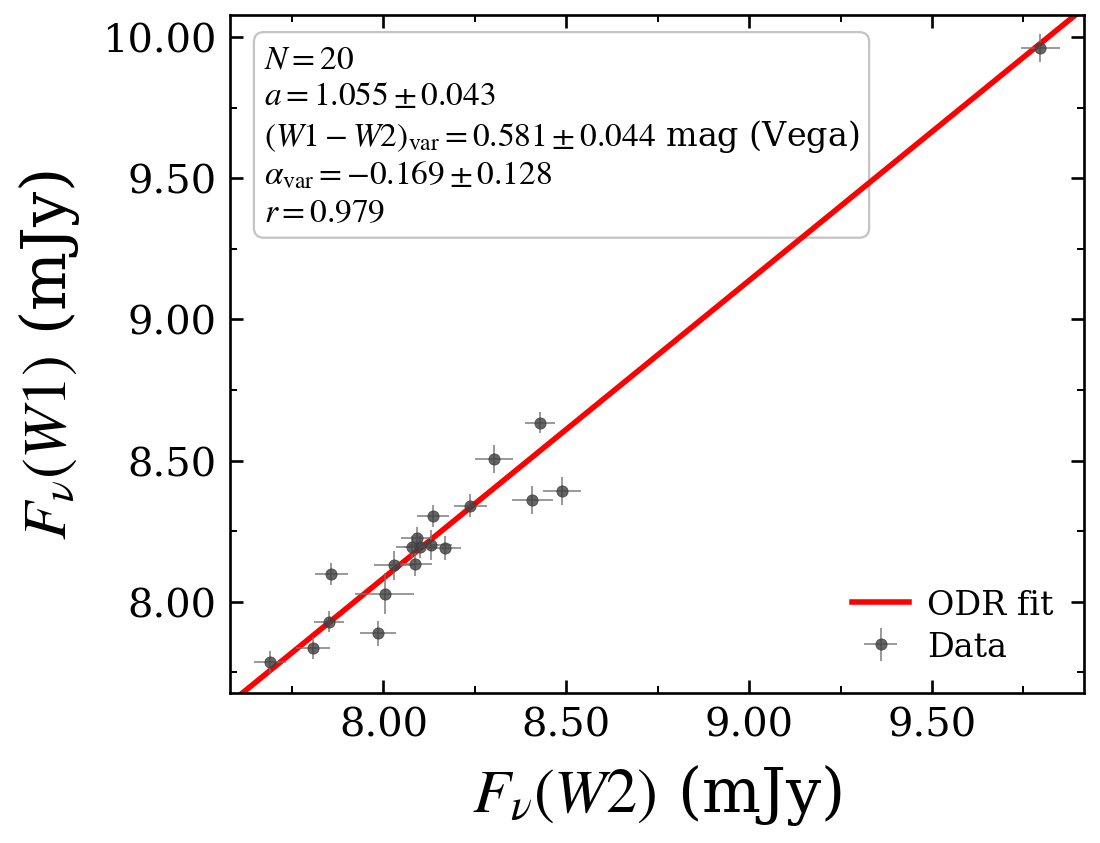}}\\[-0.5ex]
(c)
\end{minipage}

\caption{Flux--flux relations for Mrk~42.
Panel (a) shows $F_{\nu}(g)$ versus $F_{\nu}(r)$,
panel (b) shows $F_{\nu}(r)$ versus $F_{\nu}(i)$, and
panel (c) shows the epoch-averaged
$F_{\nu}(W1)$ versus $F_{\nu}(W2)$ relation.}
\label{fig:mrk42_fvg}
\end{figure*}

The optical and $W1/W2$ flux--flux relations remain close to linear
(Fig.~\ref{fig:mrk42_fvg}). Their slopes indicate that the
dominant correlated variable component preserves a broadly stable
effective spectral shape over the sampled timescales.
The corresponding variable-component spectral indices are
$\alpha_{\rm var}=0.177\pm0.085$,
$-0.353\pm0.397$, and $-0.169\pm0.128$ for the
$g-r$, $r-i$, and $W1-W2$ relations, respectively. 
The fitted flux--flux slopes and variable-component spectral
indices are listed in Table~\ref{tab:mrk42_state_optlags}, together
with the flare-state optical inter-band lags.
In the convention $F_{\nu}\propto\nu^{-\alpha}$,
the negative values of ${\rm d}\alpha/{\rm d}(-X)$
correspond to the observed bluer-when-brighter behaviour.

\begin{table}
\centering
\caption{Colour--magnitude correlations and spectral parameters
for Mrk~42.}
\label{tab:mrk42_spec}
\footnotesize
\setlength{\tabcolsep}{1.8pt}
\renewcommand{\arraystretch}{1.12}

\begin{tabular*}{\columnwidth}{@{\extracolsep{\fill}}lccc@{}}
\hline
Colour & $m$ & $r$ & $p$ \\
\hline
$g-r$ (quiet) & $0.112\pm0.085$ & 0.530 & $<10^{-6}$ \\
$g-r$ (flare) & $0.342\pm0.027$ & 0.847 & $<10^{-6}$ \\
$r-i$ (quiet) & $0.329\pm0.149$ & 0.603 & $<10^{-6}$ \\
$r-i$ (flare) & $0.800^{+0.689}_{-0.246}$ & 0.788 & $6.8{\times}10^{-3}$ \\
$W1-W2$ versus $W1$ & $0.039\pm0.036$ & 0.267 & 0.254 \\
\hline
\end{tabular*}

\par\vspace{0.8ex}

\begin{tabular*}{\columnwidth}{@{\extracolsep{\fill}}lcccc@{}}
\hline
Colour & $K$ & $\widetilde{C}$ & $\alpha_{\rm median}$ &
${\rm d}\alpha/{\rm d}(-X)$ \\
\hline
$g-r$ (quiet) & 0.319 & 0.625 & 1.961 & $-0.352\pm0.266$ \\
$g-r$ (flare) & 0.319 & 0.515 & 1.615 & $-1.073\pm0.084$ \\
$r-i$ (quiet) & 0.225 & 0.167 & 0.744 & $-1.465\pm0.662$ \\
$r-i$ (flare) & 0.225 & 0.124 & 0.554 & $-3.563^{+1.095}_{-3.068}$ \\
$W1-W2$ versus $W1$ & 0.344 & $-0.0136$ & $-0.0394$ & $-0.112\pm0.104$ \\
\hline
\end{tabular*}

\par\vspace{0.5ex}

\parbox{\columnwidth}{\footnotesize
Notes. The quantity $m$ is the York colour--magnitude slope;
$r$ and $p$ are the Pearson coefficient and associated
$p$-value; $K$ is the colour-to-spectral-index conversion factor;
$\widetilde{C}$ is the median colour; and
${\rm d}\alpha/{\rm d}(-X)$ is the spectral-index slope with
brightness, where $X=g$, $r$, or $W1$, respectively.
The WISE median colour and $\alpha_{\rm median}$ are given in
the AB system, while the fitted $W1-W2$ relation uses Vega
magnitudes. The median colours and spectral indices describe
unresolved central-region emission and are therefore not
host-subtracted intrinsic AGN quantities.
Uncertainties are derived from the 16th--84th percentile
bootstrap intervals and displayed symmetrically, except
for the flare-state $r-i$ slope and its spectral-index derivative.
}
\end{table}

\begin{table}
\centering
\caption{Flux--flux relations and flare-state optical
inter-band lags for Mrk~42.}
\label{tab:mrk42_state_optlags}
\footnotesize
\setlength{\tabcolsep}{2.2pt}
\renewcommand{\arraystretch}{1.08}

Flux--flux relations

\begin{tabular}{@{}p{0.43\columnwidth}ccc@{}}
\hline
Relation & $a$ & $r$ & $\alpha_{\rm var}$ \\
\hline

$F_{\nu}(g)$ versus $F_{\nu}(r)$
& $0.949\pm0.024$
& 0.928
& $0.177\pm0.085$ \\

$F_{\nu}(r)$ versus $F_{\nu}(i)$
& $1.076\pm0.088$
& 0.877
& $-0.353\pm0.397$ \\

$F_{\nu}(W1)$ versus $F_{\nu}(W2)$
& $1.055\pm0.043$
& 0.979
& $-0.169\pm0.128$ \\

\hline
\end{tabular}

\vspace{1.0ex}

Flare-state optical inter-band lags

\begin{tabular}{@{}lcc@{}}
\hline
Band pair & $\tau_{\rm cent}$ (d) & Response \\
\hline

$g$ versus $i$
& $3.6^{+2.2}_{-2.7}$
& $i$ follows $g$ \\

$r$ versus $i$
& $4.9^{+1.9}_{-3.4}$
& $i$ follows $r$ \\

\hline
\end{tabular}

\vspace{0.5ex}

\parbox{\columnwidth}{\footnotesize
Notes. 
The quantity $a$ is the flux--flux slope and
$\alpha_{\rm var}$ is the spectral index of the correlated
variable component ($F_{\nu}\propto\nu^{-\alpha_{\rm var}}$).
Their uncertainties are symmetric representations of the
16th--84th percentile bootstrap intervals.
The optical inter-band delays are derived from the
FR/RSS ICCF centroid distributions. The flare-state diagnostics
associated with the quoted positive $i$-band delays are shown in
Fig.~\ref{fig:app_mrk42_optlags} of
Appendix~\ref{app:lag_diagnostics}. The quiet-state lag estimates
are consistent with zero within their FR/RSS uncertainties.
}

\end{table}

The optical inter-band results provide an independent diagnostic of
the state-dependent $r-i$ relation. In the quiet state, the
$g/r/i$ delays are compatible with zero within their uncertainties.
During the flare, the centroid distributions place the $i$-band
response several days after the $g$ and $r$ variations. Although
the measured delays are only a few days, their positive sense
follows the expected wavelength ordering of continuum reprocessing
toward the longer-wavelength $i$ band. This behaviour also agrees with the steeper flare-state $r-i$
relation and indicates an additional slowly responding contribution
at the longest ZTF wavelength
(see Section~\ref{sec:disc_mrk42_flare} for more details).

\subsubsection{Mrk~493}
\label{sec:ugc10120_colours}

\begin{figure*}
\centering

\begin{minipage}{0.32\textwidth}
\centering
\includegraphics[width=\linewidth]
{\detokenize{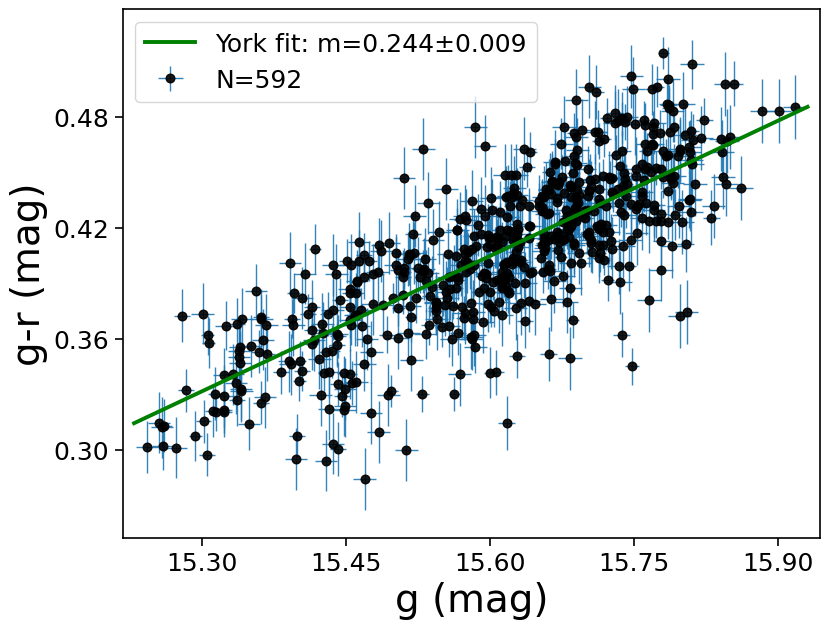}}\\[-0.5ex]
(a)
\end{minipage}
\hfill
\begin{minipage}{0.32\textwidth}
\centering
\includegraphics[width=\linewidth]
{\detokenize{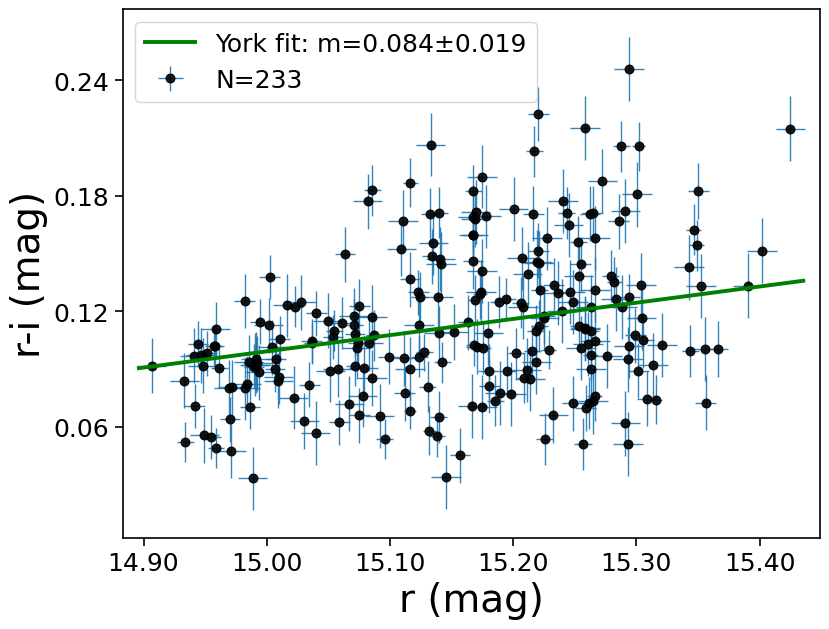}}\\[-0.5ex]
(b)
\end{minipage}
\hfill
\begin{minipage}{0.32\textwidth}
\centering
\includegraphics[width=\linewidth]
{\detokenize{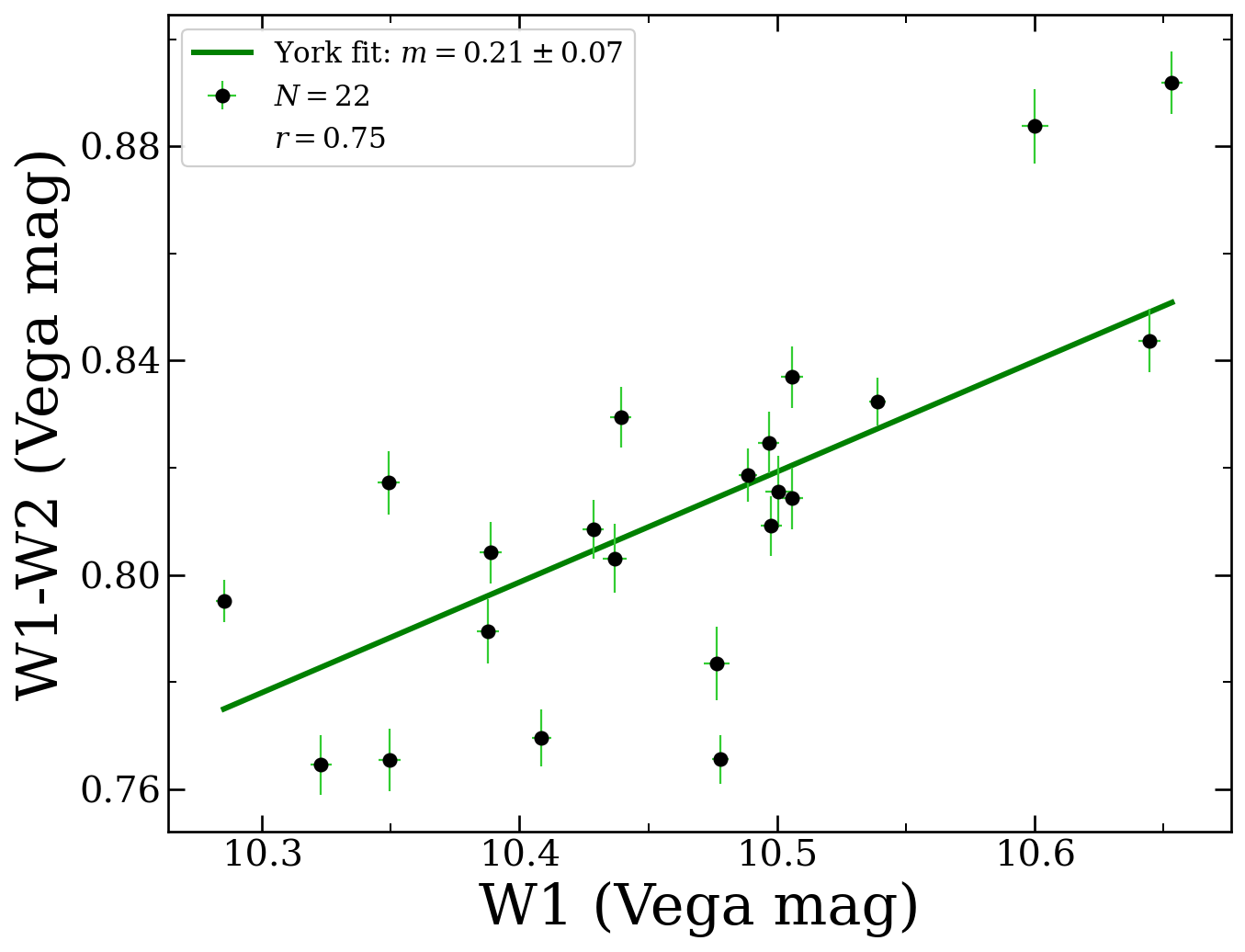}}\\[-0.5ex]
(c)
\end{minipage}

\caption{Colour--magnitude relations for Mrk~493.
Panel (a) shows $g-r$ versus $g$, panel (b) shows
$r-i$ versus $r$, and panel (c) shows
$W1-W2$ versus $W1$. The relations describe unresolved
central-region emission containing nuclear, host-galaxy, and
circumnuclear contributions.}
\label{fig:ugc10120_colors}
\end{figure*}

\begin{figure*}
\centering

\begin{minipage}{0.32\textwidth}
\centering
\includegraphics[width=\linewidth]
{\detokenize{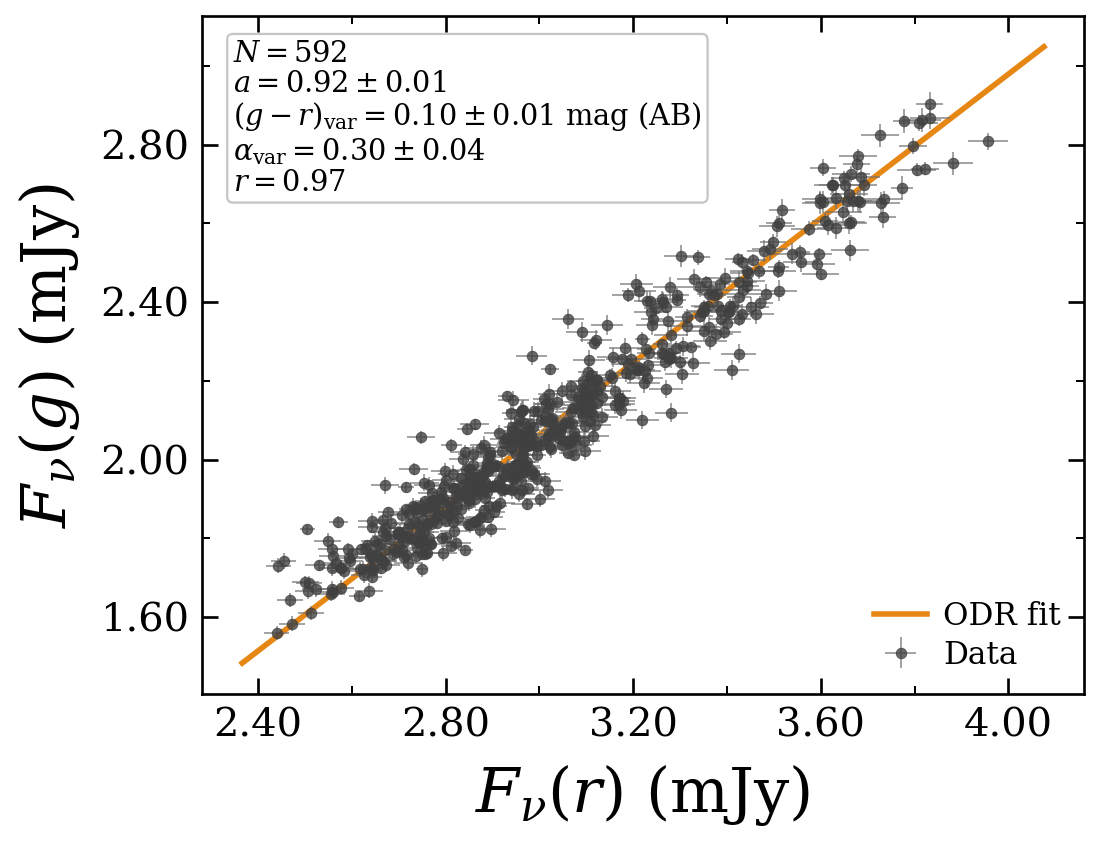}}\\[-0.5ex]
(a)
\end{minipage}
\hfill
\begin{minipage}{0.32\textwidth}
\centering
\includegraphics[width=\linewidth]
{\detokenize{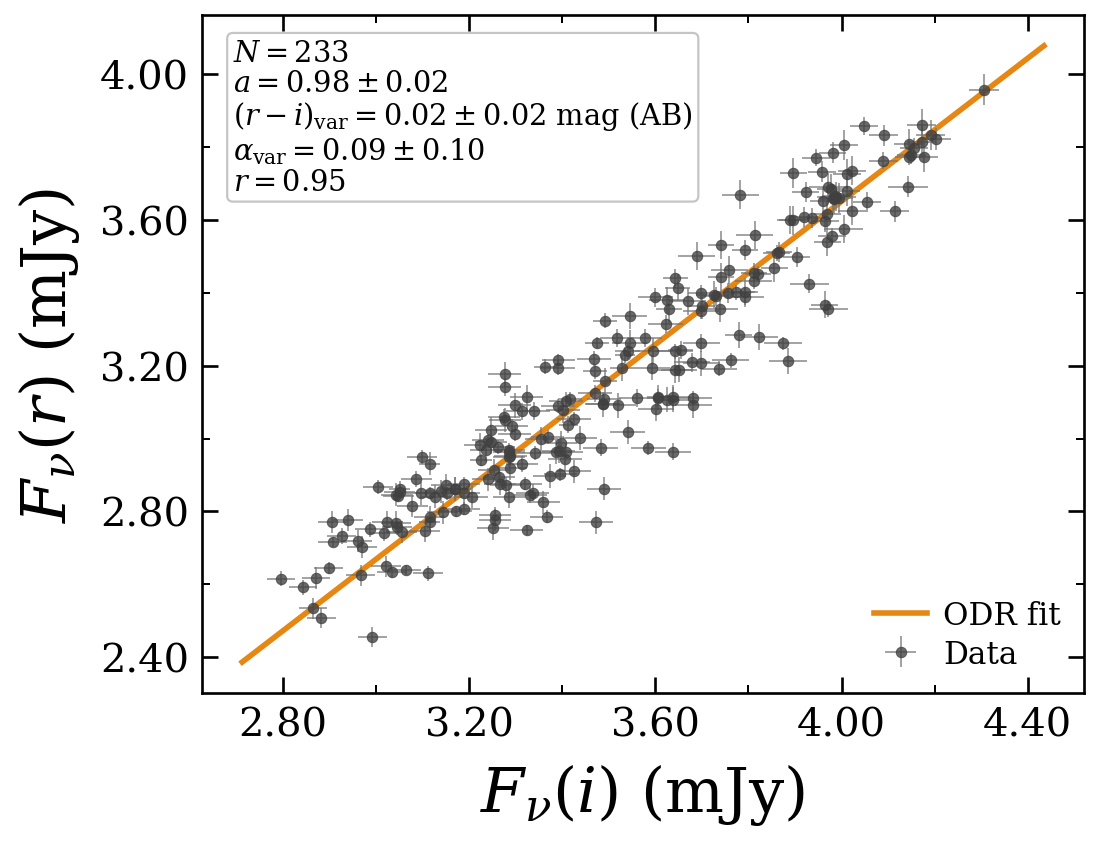}}\\[-0.5ex]
(b)
\end{minipage}
\hfill
\begin{minipage}{0.32\textwidth}
\centering
\includegraphics[width=\linewidth]
{\detokenize{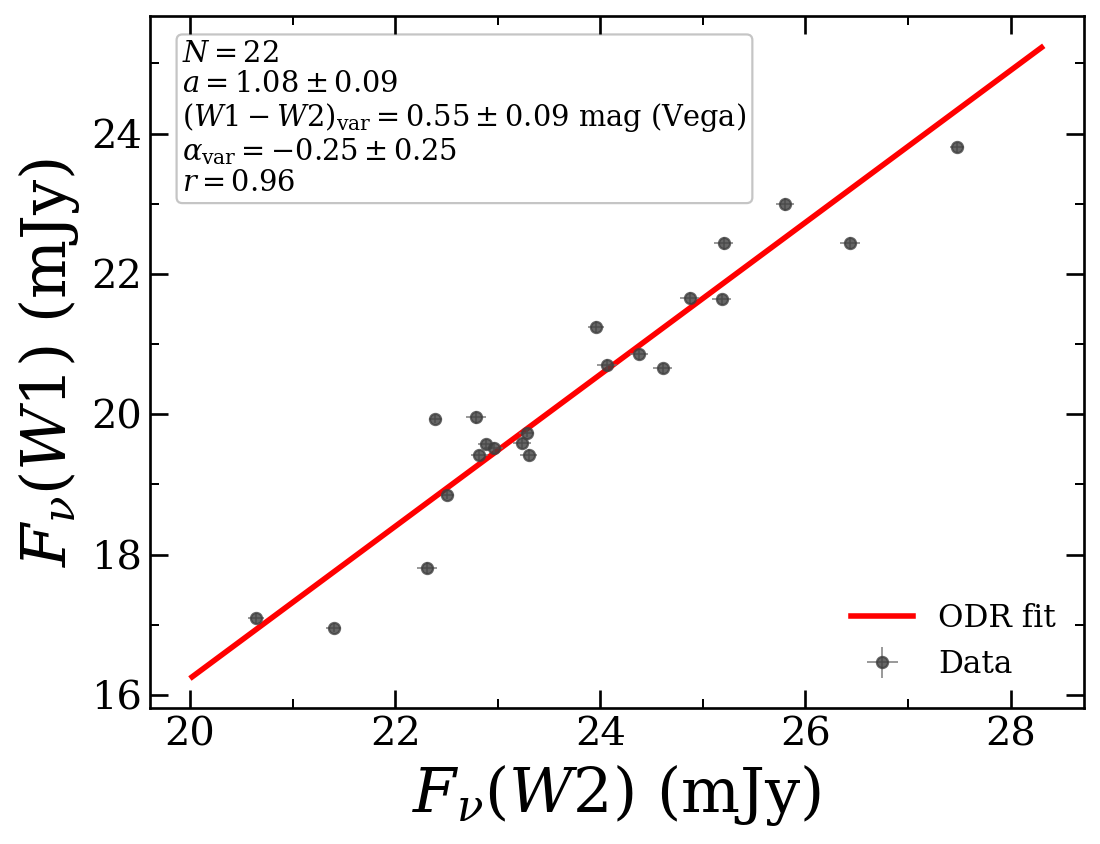}}\\[-0.5ex]
(c)
\end{minipage}

\caption{Flux--flux relations for Mrk~493.
Panel (a) shows $F_{\nu}(g)$ versus $F_{\nu}(r)$,
panel (b) shows $F_{\nu}(r)$ versus $F_{\nu}(i)$, and
panel (c) shows the epoch-averaged
$F_{\nu}(W1)$ versus $F_{\nu}(W2)$ relation.}
\label{fig:ugc10120_fvg}
\end{figure*}

Mrk~493 also shows statistically significant chromatic variability
(Fig.~\ref{fig:ugc10120_colors}). The $g-r$ versus $g$
relation has a pronounced bluer-when-brighter slope, while the
$r-i$ versus $r$ relation is weaker but remains significant. The
$W1-W2$ versus $W1$ relation shows a measurable colour response
across the epoch-averaged WISE observations. The fitted parameters
are given in Table~\ref{tab:ugc10120_ztfwise}.
The optical and WISE flux--flux relations are approximately linear
(Fig.~\ref{fig:ugc10120_fvg}), indicating broadly stable
effective colours of the dominant correlated variable components
over the sampled timescales.
The corresponding variable-component spectral indices are
$\alpha_{\rm var}=0.30\pm0.04$,
$0.09\pm0.10$, and $-0.25\pm0.25$ for the
$g-r$, $r-i$, and $W1-W2$ flux--flux relations, respectively.
The $r-i$ and $W1-W2$ variable-component spectral indices
are consistent with zero within their uncertainties.

\begin{table}
\centering
\caption{Colour--magnitude correlations and spectral parameters
for Mrk~493.}
\label{tab:ugc10120_ztfwise}
\footnotesize
\setlength{\tabcolsep}{1.8pt}
\renewcommand{\arraystretch}{1.12}

\begin{tabular}{@{}lccccccc@{}}
\hline
Colour & $m$ & $r$ & $p$ & $K$ & $\widetilde{C}$ &
\shortstack[c]{$\alpha_{\rm median}$} &
\shortstack[c]{${\rm d}\alpha/{\rm d}(-X)$} \\
\hline

\shortstack[l]{$g-r$\\versus $g$}
& \shortstack[c]{$0.24$\\
                 $\pm0.01$}
& 0.791
& $<10^{-6}$
& 0.319
& 0.41
& 1.29
& \shortstack[c]{$-0.77$\\
                 $\pm0.03$} \\

\shortstack[l]{$r-i$\\versus $r$}
& \shortstack[c]{$0.08$\\
                 $\pm0.02$}
& 0.364
& $1.1{\times}10^{-8}$
& 0.225
& 0.11
& 0.47
& \shortstack[c]{$-0.38$\\
                 $\pm0.08$} \\

\shortstack[l]{$W1-W2$\\versus $W1$}
& \shortstack[c]{$0.21$\\
                 $\pm0.07$}
& 0.753
& $5.2{\times}10^{-5}$
& 0.344
& 0.17
& 0.50
& \shortstack[c]{$-0.60$\\
                 $\pm0.19$} \\

\hline
\end{tabular}

\vspace{0.5ex}

\parbox{\columnwidth}{\footnotesize
Notes. The notation follows Table~\ref{tab:mrk42_spec}.
The WISE median colour and $\alpha_{\rm median}$ are reported in
the AB system, while the fitted $W1-W2$ relation uses native Vega
magnitudes. The colour-derived quantities describe the unresolved
central region and are therefore not host-subtracted intrinsic AGN
values.
The fitted-slope and spectral-index uncertainties are based on
the 16th--84th percentile bootstrap intervals, displayed as
symmetric averages of the upper and lower errors.
}
\end{table}

Because the ZTF and WISE measurements contain unresolved host and
circumnuclear emission, the colour--magnitude slopes describe the
combined central-region signal. A relatively red, slowly varying host contribution can dilute the
fractional nuclear variability and enhance the apparent
bluer-when-brighter trend as the variable nuclear contribution
increases. In the WISE bands, unresolved host and circumnuclear
emission can similarly reduce the amplitude of the observed
$W1-W2$ changes. The median colours and colour-derived spectral
indices are therefore used as observed central-region diagnostics,
while the flux--flux slopes characterise the correlated variable
components.

\subsection{Optical--infrared reverberation lags and hot-dust radii}
\label{subsec:results_reverberation}

The primary optical--infrared delays were obtained from the
FR/RSS ICCF centroid distributions using the
accretion-disc-corrected $W1$ and $W2$ light curves. The
measurements are summarised jointly in
Table~\ref{tab:combined_dust_lags}. Positive lags are recovered in
both bands for both galaxies.
For Mrk~42, the 5000-realisation FR/RSS analysis gives
$\tau_{\rm obs}(W1)=54^{+13}_{-33}$~d and
$\tau_{\rm obs}(W2)=62^{+7}_{-35}$~d.
For Mrk~493, the 5000-realisation FR/RSS analysis gives
$\tau_{\rm obs}(W1)=94^{+18}_{-21}$~d and
$\tau_{\rm obs}(W2)=111^{+26}_{-24}$~d.
Applying the redshift/wavelength correction of
Equation~\ref{eq:tau_corr} gives response-weighted dust radii of
$0.045^{+0.011}_{-0.028}$ and
$0.052^{+0.006}_{-0.029}$~pc for the Mrk~42 $W1$ and $W2$
responses, respectively, and
$0.078^{+0.015}_{-0.017}$ and
$0.092^{+0.021}_{-0.020}$~pc for Mrk~493.

The pyZDCF calculations provide independent checks on the
optical--infrared correlations. Because the approximately six-month
WISE cadence produces broad correlation constraints, the pyZDCF
results are interpreted only as qualitative consistency checks
rather than as independent lag measurements. For Mrk~42, the $W1$
and $W2$ pyZDCF functions show broad plateaus spanning near-zero
and positive lags, with formal maxima at approximately 31~d in both
bands. These maxima lie within the lower portions of the
corresponding FR/RSS ICCF intervals, while descriptive
correlation-weighted centroids are shifted toward shorter lags by
the near-zero bins.

For Mrk~493, the pyZDCF calculations also recover broad
positive-lag features compatible with the fiducial ICCF solutions.
The extended ATLAS+ZTF driver gives
$\tau_{\rm obs}(W1)=93^{+16}_{-17}$~d and
$\tau_{\rm obs}(W2)=111^{+24}_{-21}$~d, consistent with the
fiducial ZTF-only values of
$94^{+18}_{-21}$ and $111^{+26}_{-24}$~d, respectively
(Fig.~\ref{fig:app_mrk493_iccf}).
 In both objects, the reported
reverberation lags are therefore defined by the FR/RSS ICCF
centroid distributions. 
The complete lag-diagnostic figures are collected in
Appendix~\ref{app:lag_diagnostics}, while additional implementation
details are given in Appendix~\ref{app:lag_method_details}.

Sensitivity tests with $s_{\rm AD}=0.1$ and $s_{\rm AD}=0$
yielded centroid-lag distributions that overlap the fiducial
$s_{\rm AD}=1/3$ intervals for both objects and both WISE bands.
As a separate robustness test for Mrk~42, we repeated the
centroid-based ICCF/FR/RSS analysis after subtracting the fitted
47.12-d residual component from the optical driver
(see Section~\ref{sec:mrk42_flare_qpo} for details).
We obtained consistent results, indicating that the inferred
optical--infrared delays are insensitive to the 47.12-d
intra-flare optical structure
(Fig.~\ref{fig:mrk42_frrss_robustness};
Appendix~\ref{app:lag_diagnostics}).

\begin{table}
\centering
\caption{Optical--infrared reverberation measurements and
redshift/wavelength-corrected hot-dust radii.}
\label{tab:combined_dust_lags}
\setlength{\tabcolsep}{2.5pt}
\renewcommand{\arraystretch}{1.10}

\begin{tabular}{@{}llccc@{}}
\hline
Object & Band &
$\tau_{\rm obs}$ &
$\tau_{\rm corr}$ &
$R_{\rm dust}$ \\
& & (d) & (d) & (pc) \\
\hline

Mrk~42
& $W1$
& $54^{+13}_{-33}$
& $53^{+13}_{-33}$
& $0.045^{+0.011}_{-0.028}$ \\

Mrk~42
& $W2$
& $62^{+7}_{-35}$
& $62^{+7}_{-35}$
& $0.052^{+0.006}_{-0.029}$ \\

Mrk~493
& $W1$
& $94^{+18}_{-21}$
& $93^{+18}_{-21}$
& $0.078^{+0.015}_{-0.017}$ \\

Mrk~493
& $W2$
& $111^{+26}_{-24}$
& $110^{+25}_{-24}$
& $0.092^{+0.021}_{-0.020}$ \\

\hline
\end{tabular}

\vspace{0.5ex}

\parbox{\columnwidth}{
Notes. The listed observed-frame lags are medians of the FR/RSS ICCF
centroid distributions. 
The corrected lags were calculated as
$\tau_{\rm corr}=\tau_{\rm obs}(1+z)^{-0.38}$ following
\citet{Mandal2024DustTorus}, and the response-weighted dust radii
were obtained from $R_{\rm dust}=c\tau_{\rm corr}$.
}
\end{table}

\subsection{Contextual BLR and hot-dust scales}
\label{subsec:results_radial_context}

The source parameters adopted for the radial-scale comparisons are
collected in Table~\ref{tab:combined_source_parameters}. For
Mrk~42, the AGN continuum luminosity is derived from the
host--AGN decomposition of the SDSS spectrum. For Mrk~493, the
host-corrected luminosity and direct H$\beta$ reverberation lag are
taken from the SEAMBH analysis of \citet{Du2015SEAMBH4}.

\begin{table}
\centering
\caption{Source parameters used in the BLR and hot-dust
comparisons.}
\label{tab:combined_source_parameters}
\setlength{\tabcolsep}{1.4pt}
\renewcommand{\arraystretch}{1.04}

\begin{tabular}{@{}p{0.54\columnwidth}
                    p{0.28\columnwidth}
                    p{0.10\columnwidth}@{}}
\hline
Quantity & Value & Ref. \\
\hline

\multicolumn{3}{@{}l}{\textit{Mrk~42}} \\[0.2ex]

$z_{\rm adopted}$
& 0.024251
& [1] \\

FWHM(H$\beta$)
& 940~km~s$^{-1}$
& [2] \\

FWHM(H$\beta$)
& 973~km~s$^{-1}$
& [3] \\

$\log(M_{\bullet}/M_{\odot})$
& 6.37
& [3] \\

$M_{\bullet}$
& $(2.5^{+0.7}_{-0.6})\times10^{6}\,M_{\odot}$
& [4] \\

$\log\lambda L_{\lambda}(5100\,\mbox{\AA})_{\rm tot}$
& 42.84
& [2] \\

$\log\lambda L_{\lambda}(5100\,\mbox{\AA})_{\rm tot}$
& 42.88
& [3] \\

$\log\lambda L_{\lambda}(5100\,\mbox{\AA})_{\rm tot}$
& $42.85$
& [TW] \\

$\log\lambda L_{\lambda}(5100\,\mbox{\AA})_{\rm AGN}$,
fiducial
& $42.55\pm0.07$
& [TW] \\

$\log\lambda L_{\lambda}(5100\,\mbox{\AA})_{\rm AGN}$,
broad mask
& $42.62$
& [TW] \\

\hline

\multicolumn{3}{@{}l}{\textit{Mrk~493}} \\[0.2ex]

$z_{\rm SEAMBH}$
& 0.0313
& [5] \\

$z_{\rm adopted}$
& 0.031503
& [1] \\

Environment
& isolated galaxy
& [6] \\

Nearby companions
& 2
& [6] \\

$\tau_{{\rm H}\beta,\rm rest}$
& $11.6^{+1.2}_{-2.6}$~d
& [5] \\

FWHM(H$\beta$)
& $778\pm12$~km~s$^{-1}$
& [5] \\

$\log(M_{\bullet}/M_{\odot})$
& 6.14
& [5] \\

$\log\dot{M}$
& 1.88
& [5] \\

$\log\lambda L_{\lambda}(5100\,\mbox{\AA})_{\rm AGN}$
& $43.11\pm0.08$
& [5] \\

$\log L_{{\rm H}\beta}$
& $41.35\pm0.05$
& [5] \\

EW(H$\beta$)
& $87.4\pm18.1$~\AA
& [5] \\

\hline
\end{tabular}

\vspace{0.5ex}

\parbox{\columnwidth}{
Notes. References: [1] NED; [2] \citet{Sani2010};
[3] \citet{Du2014BLRNLR}; [4] \citet{Hennig2018};
[5] \citet{Du2015SEAMBH4}; [6] \citet{Pulatova2015};
[TW] this work. Logarithmic luminosities are given in
${\rm erg\,s^{-1}}$.
}
\end{table}

For a common luminosity-based BLR reference, we used the
average-scheme total-sample $R_{\rm BLR}$--$L_{5100}$ relation
given by equation~(7) of \citet{WangWoo2024}. The luminosity
uncertainties, calibration-coefficient uncertainties, and the
$0.18\pm0.02$~dex intrinsic scatter of the relation were propagated
through 50\,000 Monte Carlo realisations. As a calibration
sensitivity test, using the alternative total-sample relation of
\citet{Woo2024SAMP} changes the inferred $\log R_{\rm BLR}$ values
by only 0.05--0.06~dex, well below the intrinsic scatter of either
relation. The fiducial calibration gives central contextual BLR
radii of 0.005~pc for Mrk~42 and 0.008~pc for Mrk~493. The median
relation-based dust-to-BLR ratios, quoted here as approximate
integer comparisons, are 8 and 9 for the Mrk~42 $W1$ and $W2$
responses, respectively, and 9 and 11 for Mrk~493.

For Mrk~493, the direct rest-frame H$\beta$ lag of
$11.6^{+1.2}_{-2.6}$~d corresponds to a measured BLR radius of
$0.010^{+0.001}_{-0.002}$~pc. Using this direct scale, the
corresponding dust-to-BLR ratios are approximately 8 for $W1$
and 10 for $W2$. These approximate ratios are consistent with
the ensemble dust-to-BLR scales reported by
\citet{Chen2023,Mandal2024DustTorus,Ayubinia2025}.

For the direct comparison with the dust size--luminosity relations,
we calculated
$L_{\rm bol}=9.26\,\lambda L_{\lambda}(5100\,\text{\AA})$,
following the luminosity convention adopted by
\citet{Mandal2024DustTorus}. This gives
\[
\log(L_{\rm bol}/{\rm erg\,s^{-1}})
=
43.52\pm0.07
\]
for Mrk~42 and
\[
\log(L_{\rm bol}/{\rm erg\,s^{-1}})
=
44.08\pm0.08
\]
for Mrk~493.

\begin{figure}
\centering
\includegraphics[width=\columnwidth]
{\detokenize{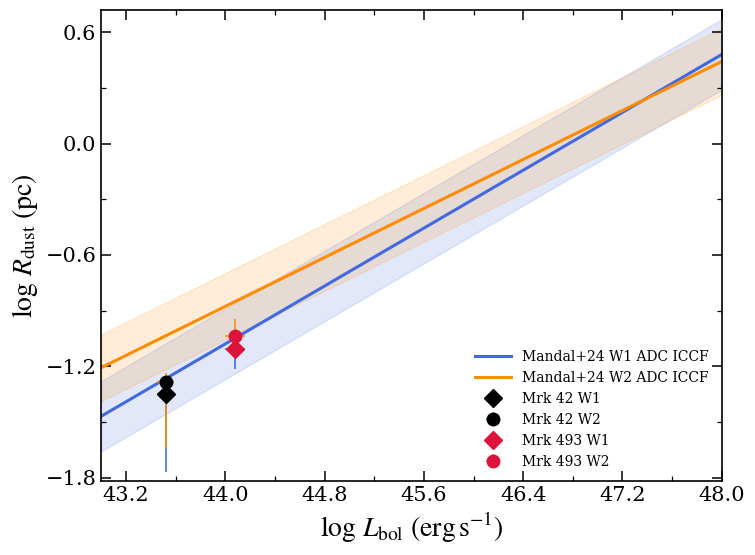}}

\caption{Mrk~42 and Mrk~493 on the
$R_{\rm dust}$--$L_{\rm bol}$ plane.
The plotted measurements use dust radii calculated from the
redshift/wavelength-corrected lags defined by
Equation~\ref{eq:tau_corr}, with
$R_{\rm dust}=c\,\tau_{\rm corr}$ as given by
Equation~\ref{eq:rdust}.
Bolometric luminosities were calculated using
$L_{\rm bol}=9.26\,\lambda L_{\lambda}(5100\,\mbox{\AA})$.
The curves and intrinsic-scatter regions show the
accretion-disc-corrected $W1$ and $W2$ size--luminosity relations
of \citet{Mandal2024DustTorus}.}
\label{fig:both_mandal_plane}
\end{figure}

The central $W1$ and $W2$ radii predicted by the
accretion-disc-corrected relations of
\citet{Mandal2024DustTorus} are approximately 0.054 and
0.092~pc for Mrk~42, and 0.089 and 0.141~pc for Mrk~493.
The corrected $W1$ radii lie approximately 0.08 and 0.06~dex
below the corresponding central relations for Mrk~42 and Mrk~493,
respectively. The $W2$ offsets are approximately 0.25 and
0.19~dex.
Considering the measurement uncertainties and the intrinsic
scatter of 0.19 and 0.18~dex in the $W1$ and $W2$ relations,
respectively, the two nuclei occupy the expected low-luminosity
region of the hot-dust size--luminosity plane
(Fig.~\ref{fig:both_mandal_plane}).

\subsection{Mrk~42 flare substructure}
\label{sec:mrk42_flare_qpo}

\begin{figure*}
\centering

\begin{minipage}{0.48\textwidth}
\centering
\includegraphics[width=\linewidth]
{\detokenize{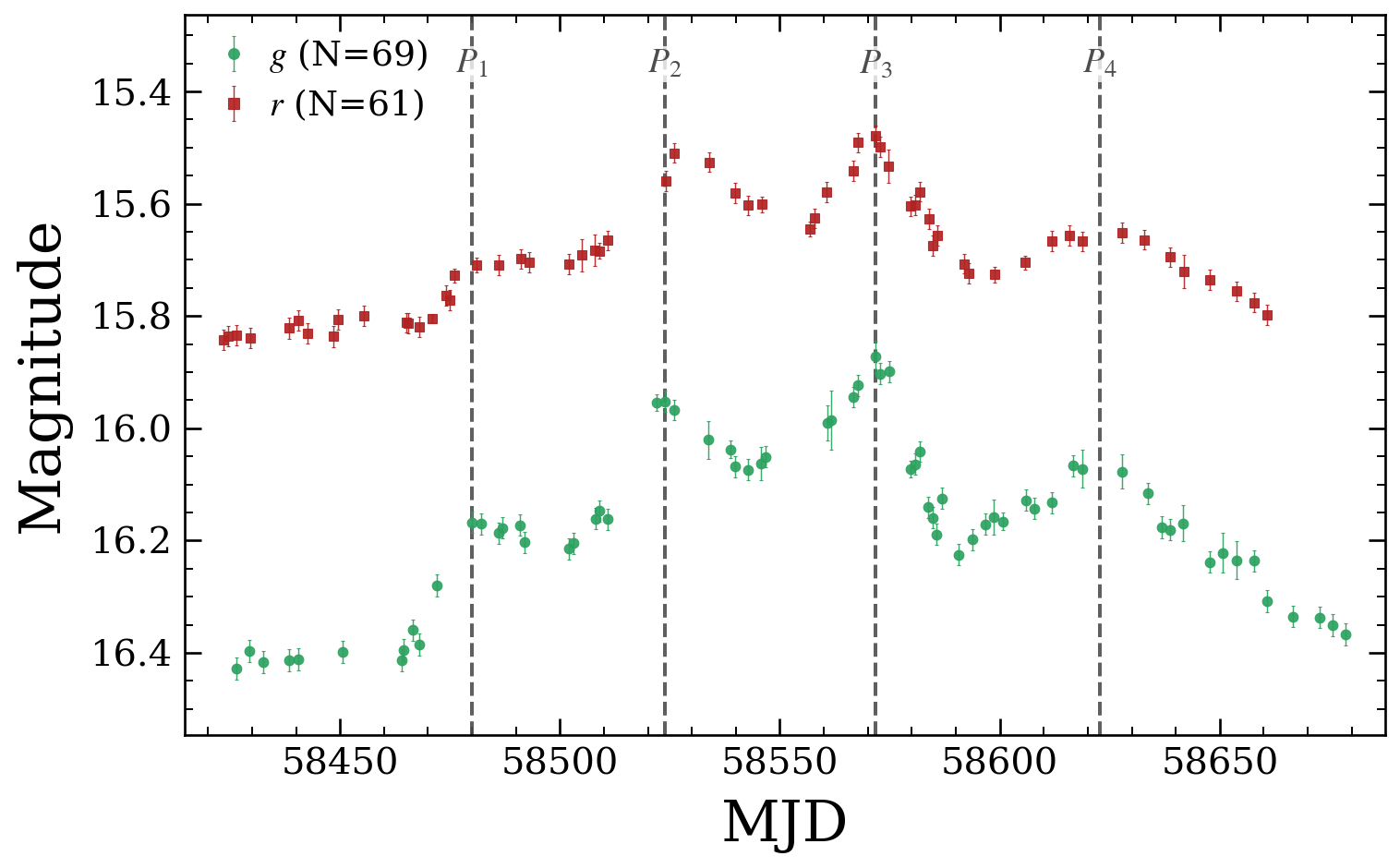}}\\[-0.5ex]
\textbf{(a)}
\end{minipage}
\hfill
\begin{minipage}{0.48\textwidth}
\centering
\includegraphics[width=\linewidth]
{\detokenize{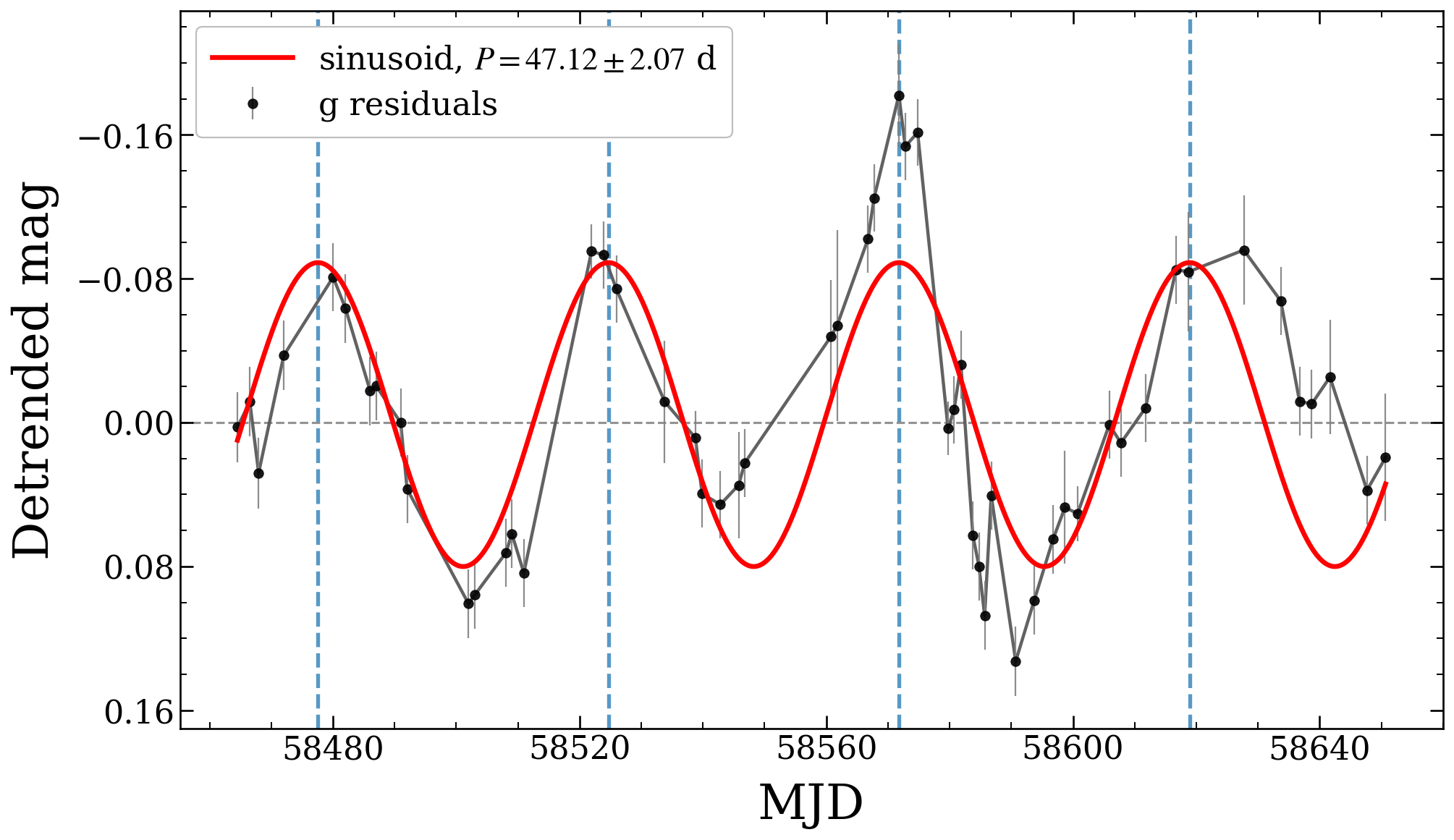}}\\[-0.5ex]
\textbf{(b)}
\end{minipage}

\caption{
Flare morphology and characteristic intra-flare substructure in
Mrk~42. Panel (a) shows the ZTF $g$- and $r$-band light curves
and marks the four broad flare maxima. Panel (b) shows the
detrended $g$-band residuals together with a sinusoidal guide at
the dominant $g$-band GLS timescale.
}
\label{fig:mrk42_flare_morphology}
\end{figure*}

\begin{figure}
\centering
\includegraphics[width=\columnwidth]
{\detokenize{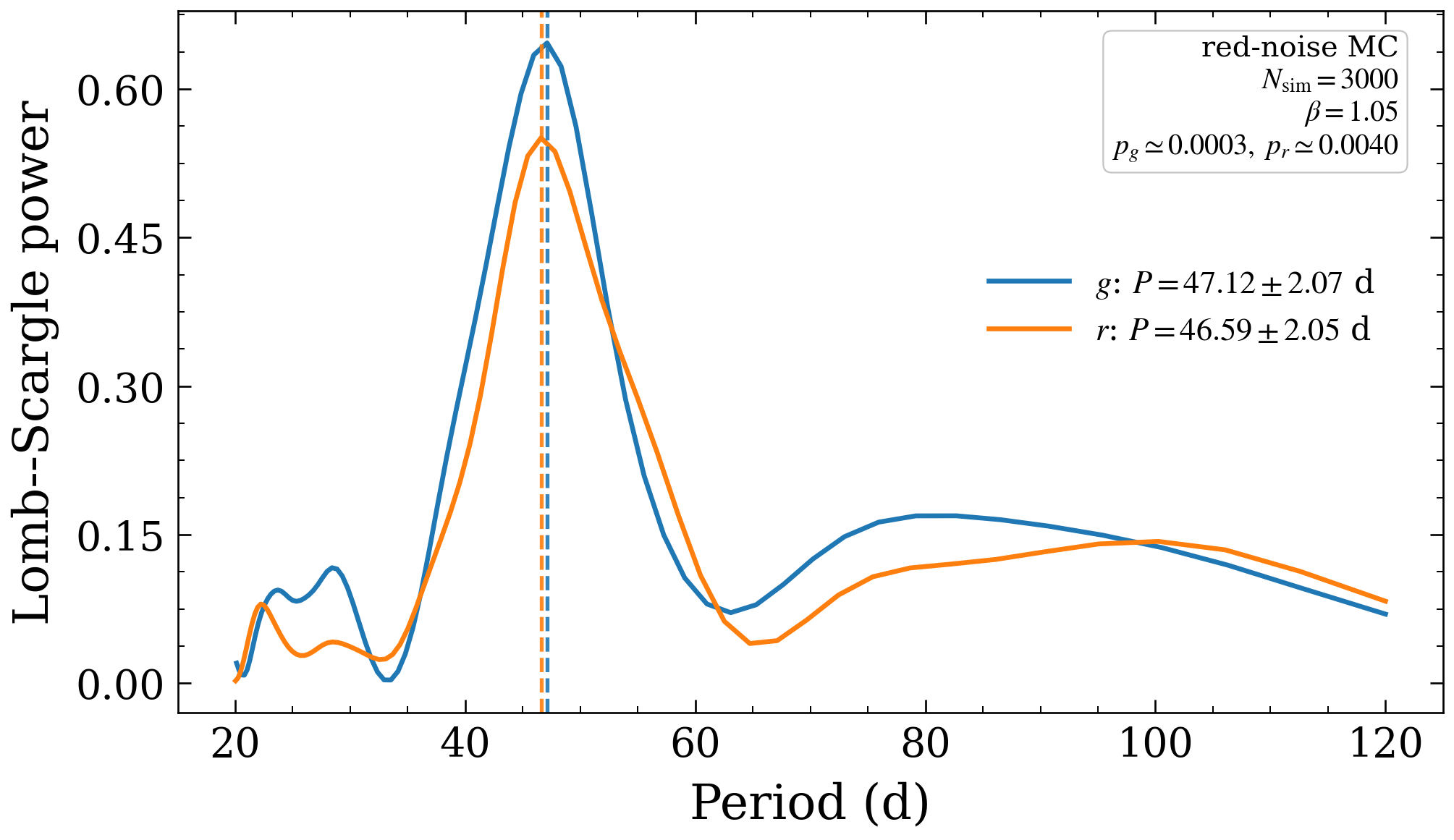}}

\caption{
Generalised Lomb--Scargle periodograms of the detrended
$g$- and $r$-band residuals of the Mrk~42 flare.
The dominant GLS features occur at 47.12~d in $g$ and
46.59~d in $r$.
}
\label{fig:mrk42_flare_periodogram}
\end{figure}

The prominent optical flare of Mrk~42 exhibits four broad maxima
superposed on an asymmetric envelope
(Fig.~\ref{fig:mrk42_flare_morphology}). After detrending the broad flare evolution, the dominant GLS
features occur at 47.12~d in $g$ and 46.59~d in $r$
(Fig.~\ref{fig:mrk42_flare_periodogram}).

The four maxima define successive peak-to-peak intervals of
 $43.9\pm1.4$, $47.8\pm1.1$, and
$50.9\pm1.1$~d. A linear fit to the four peak epochs gives a
characteristic spacing of $47.32\pm0.43$~d, with
$\chi_{\nu}^{2}=8.1$. The large fit statistic indicates that a
strictly uniform sequence does not adequately describe the four
peak epochs. The fitted spacing is therefore used as a descriptive
timescale of the intra-flare structure.

The three measured intervals increase monotonically across the
flare. This apparent lengthening is treated as a descriptive
feature because only three intervals are available and adjacent
intervals share one peak epoch.

The Gaussian-process comparison gives a characteristic stochastic
timescale of 45.1~d. The stochastic continuum model has the lower
BIC, and inclusion of the damped SHO component gives
$\Delta{\rm BIC}=+2.0$. The model comparison therefore provides
no statistical preference for the additional structured covariance
component. Together, the GLS, peak-timing, and Gaussian-process
diagnostics identify a characteristic 45--47~d intra-flare
structure within the broader optical event. The principal
flare-substructure diagnostics are summarised in
Table~\ref{tab:mrk42_flare_period}.

\begin{table}
\centering
\caption{Summary of the Mrk~42 flare-substructure analysis.}
\label{tab:mrk42_flare_period}
\footnotesize
\setlength{\tabcolsep}{3pt}
\renewcommand{\arraystretch}{1.08}

\begin{tabular}{@{}p{0.62\columnwidth}p{0.32\columnwidth}@{}}
\hline
Quantity & Value \\
\hline

Dominant GLS timescale in $g$
& 47.12~d \\

Dominant GLS timescale in $r$
& 46.59~d \\

Characteristic spacing from peak-time fit
& $47.32\pm0.43$~d \\

Reduced $\chi^{2}$ of peak-time fit
& 8.1 \\

Peak spacing, $\Delta t_{12}$
&  $43.9\pm1.4$~d \\

Peak spacing, $\Delta t_{23}$
& $47.8\pm1.1$~d \\

Peak spacing, $\Delta t_{34}$
& $50.9\pm1.1$~d \\

GP characteristic timescale
& 45.1~d \\

Lower-BIC GP model
& stochastic continuum \\

$\Delta{\rm BIC}$ (SHO--continuum)
& $+2.0$ \\

Preferred flare-envelope model
& asymmetric Gaussian \\

\hline
\end{tabular}

\vspace{0.5ex}

\parbox{\columnwidth}{\footnotesize
Note. The peak-time fit provides a descriptive characteristic
spacing. The large reduced $\chi^{2}$ reflects the non-uniform
spacing of the four peak epochs. The three successive intervals
increase monotonically, while their limited number and shared
peak-time uncertainties restrict a quantitative measurement of
the apparent trend.
}
\end{table}

\subsection{Supplementary time-domain constraints}
\label{subsec:results_supporting}

\subsubsection{IAC80 intraday monitoring of Mrk~42}
\label{sec:mrk42_idv_results}

The three IAC80 monitoring runs provide a supplementary constraint
on hour-scale optical variability during the 2025 epochs. The corresponding differential light curves are shown in
Fig.~\ref{fig:mrk42_idv} of Appendix~\ref{app:iac80}. All power-enhanced $F$ statistics
remain below the corresponding critical values at
$\alpha=0.01$ (Table~\ref{tab:mrk42_idv_stats}). The resulting
$p$-values are 0.235, 0.952, and 0.9996 for the three consecutive
nights. No significant optical intraday variability is detected
during these observations.

\begin{table}
\centering
\caption{Power-enhanced $F$-test results for Mrk~42.}
\label{tab:mrk42_idv_stats}
\footnotesize
\setlength{\tabcolsep}{4pt}
\renewcommand{\arraystretch}{1.08}

\begin{tabular}{@{}lcccc@{}}
\hline
Night & $F$ & $F_{\rm crit}$ & $p$-value & Class \\
\hline
2025-05-20/21 & 1.094 & 1.339 & 0.235 & NV \\
2025-05-21/22 & 0.811 & 1.318 & 0.952 & NV \\
2025-05-22/23 & 0.614 & 1.369 & 0.9996 & NV \\
\hline
\end{tabular}

\vspace{0.5ex}

\parbox{\columnwidth}{\footnotesize
\textit{Note.} NV denotes a non-variable classification.
}
\end{table}

\subsubsection{Swift UV/X-ray context for Mrk~493}
\label{sec:ugc10120_swift_results}

\begin{figure}
\centering
\includegraphics[width=\columnwidth]
{\detokenize{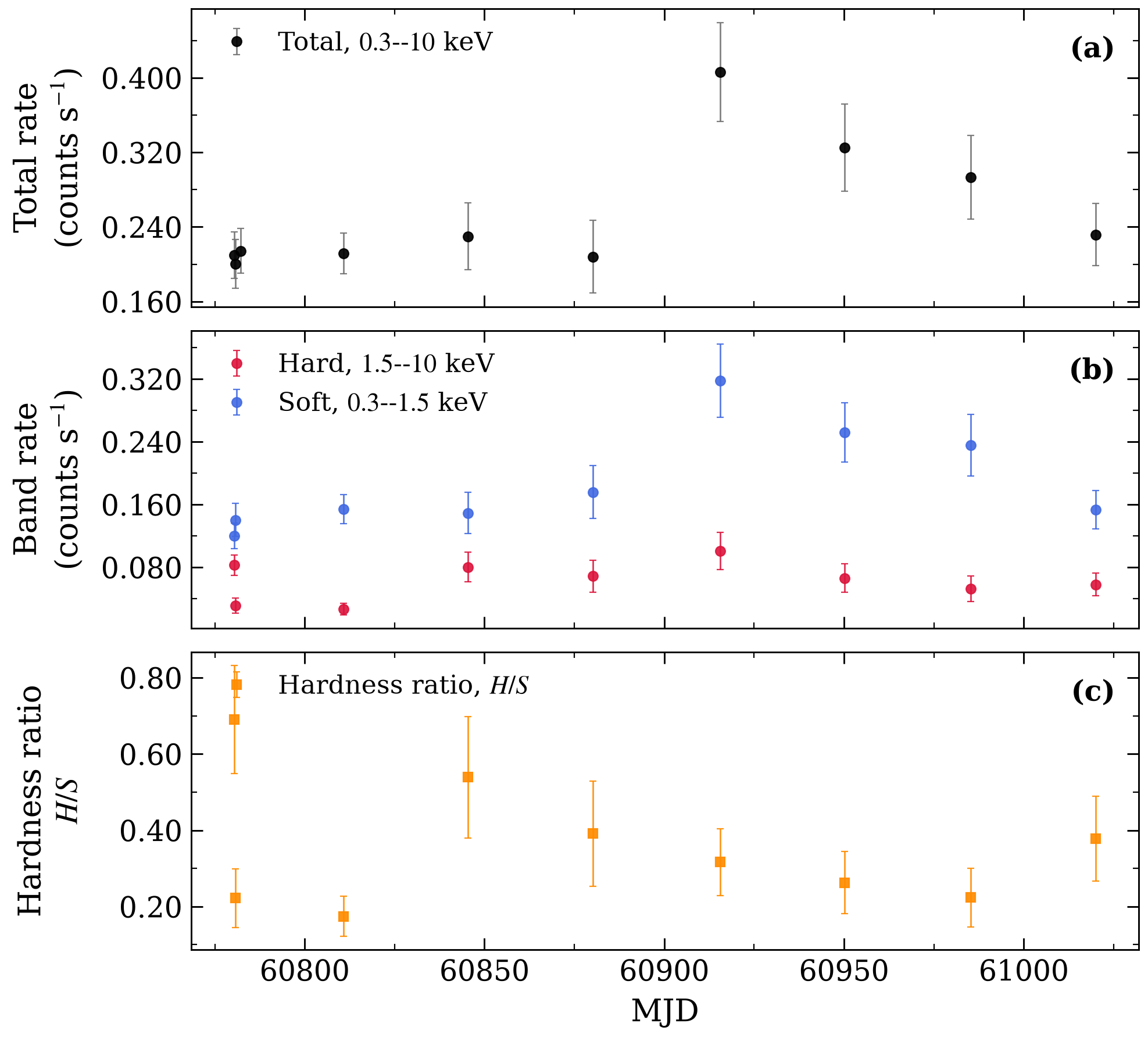}}

\caption{Epoch-binned Swift/XRT light curves of Mrk~493
during the dense monitoring interval.
The top panel shows the total 0.3--10~keV count rate,
the middle panel shows the soft 0.3--1.5~keV and hard
1.5--10~keV count rates, and the bottom panel shows the
hard-to-soft count-rate ratio,
$H/S=R_{\rm hard}/R_{\rm soft}$.
}
\label{fig:ugc10120_xrt_epochs}
\end{figure}

The Swift observations provide a complementary view of
the multiwavelength activity of Mrk~493. During the dense
monitoring interval at
$\mathrm{MJD}\simeq60780$--61020, the epoch-binned
0.3--10~keV XRT light curve contains a statistically significant
interval of enhanced X-ray count rate. The inverse-variance-
weighted baseline and enhanced-state rates are
$0.201\pm0.010$ and
$0.310\pm0.026$~counts~s$^{-1}$,
respectively, corresponding
to a difference of $3.9\sigma$. The strongest enhancement occurs
during the same broad active phase as the UV and optical brightening
(Fig.~\ref{fig:ugc10120_xrt_epochs}).

\begin{figure}
\centering
\includegraphics[width=\columnwidth]
{\detokenize{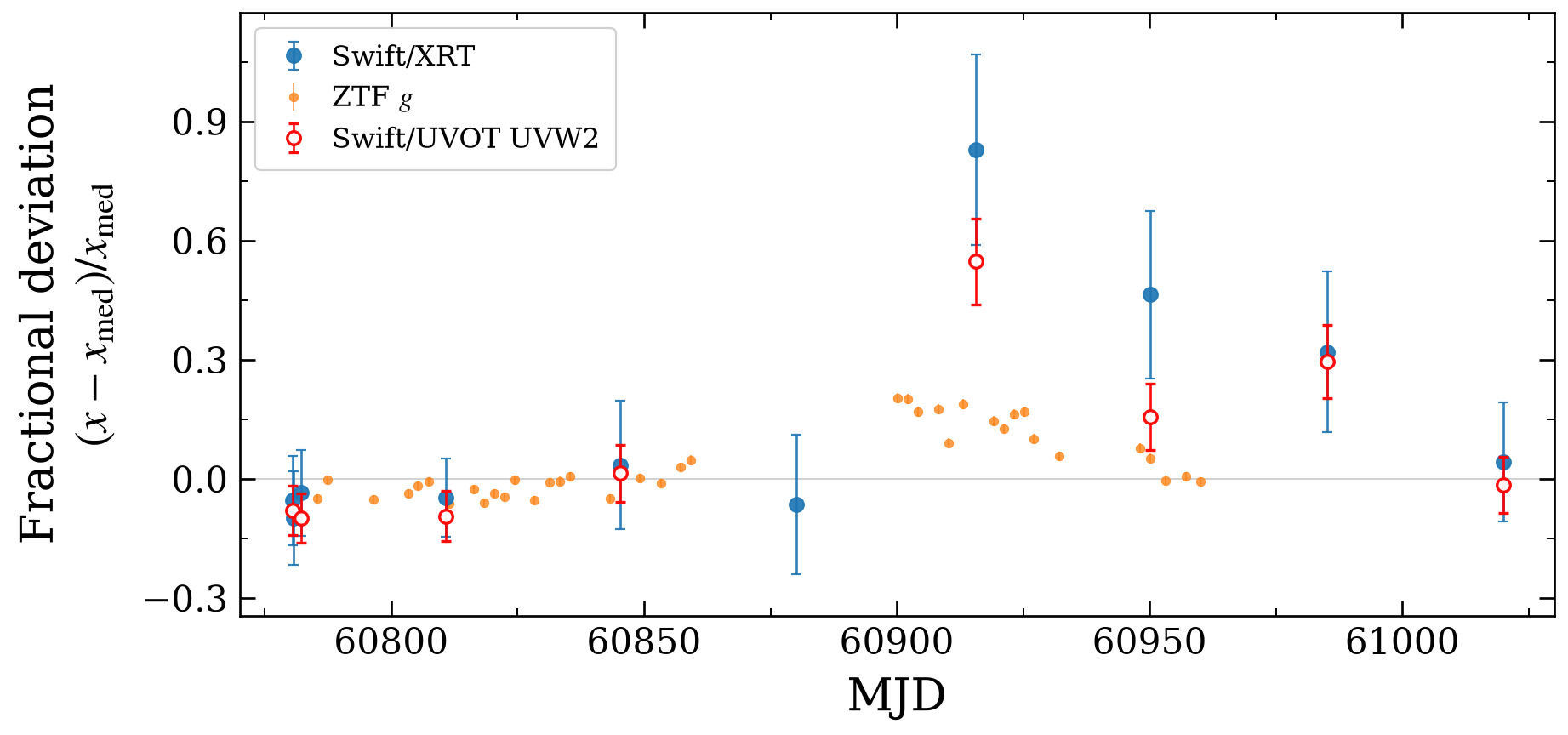}}

\caption{Fractional-deviation comparison for Mrk~493 during the
dense Swift monitoring interval. ZTF $g$-band fluxes,
UVOT UVW2 fluxes, and epoch-binned XRT 0.3--10~keV count rates
are shown relative to the median level of each series. No temporal
shifts have been applied.}
\label{fig:ugc10120_xray_uv_opt_overlay}
\end{figure}

The available cadence and temporal coverage are insufficient for
a constrained X-ray-to-UV or X-ray-to-optical lag measurement.
The fractional-deviation comparison shows broad activity in the
ZTF $g$, UVOT UVW2, and XRT series across overlapping epochs
(Fig.~\ref{fig:ugc10120_xray_uv_opt_overlay}). These measurements
are used as supplementary evidence for multi-band activity. The
optical--infrared reverberation analysis is based on the ZTF and
WISE/NEOWISE light curves.

\section{Discussion}
\label{sec:discussion}

\subsection{Hot-dust reverberation and radial context}
\label{sec:disc_dust_context}

The optical--infrared variability of Mrk~42 and Mrk~493 supports
a common interpretation in which changes in the central
optical/UV continuum are followed by a delayed response from
circumnuclear dust. The smoother epoch-averaged $W1$ and $W2$
light curves and the positive FR/RSS centroid delays place the
responding dust at sub-parsec radii in both nuclei.

This interpretation applies specifically to the short-wavelength
infrared emission sampled by the 3.4 and 4.6~$\mu$m WISE bands,
which are sensitive mainly to hot dust near the inner boundary of
the dusty structure. They probe a different component from the
extended polar emission that can dominate beyond approximately
8--10~$\mu$m \citep{Honig2013,Asmus2019}. The measured $W1/W2$ delays therefore constrain
response-weighted scales of the hot inner dust. The full radial
extent of the infrared-emitting region is not sampled by these
bands.

Mrk~493 has longer $W1$ and $W2$ delays and larger dust-response
radii than Mrk~42, in line with its higher continuum luminosity.
Both objects occupy the low-luminosity extension of the
accretion-disc-corrected $R_{\rm dust}$--$L_{\rm bol}$ relations of
\citet{Mandal2024DustTorus}, with their redshift/wavelength-corrected
radii lying within the intrinsic scatter of the published relations.
After applying the redshift/wavelength correction, both sources
remain below the central $R_{\rm dust}$--$L_{\rm bol}$ relations of
\citet{Mandal2024DustTorus}. These authors also found decreasing
torus-size residuals with increasing Eddington ratio, corresponding
to relatively smaller dust radii at higher accretion rates. Our
two objects do not by themselves provide an independent test of
this population trend.

The inferred dust-to-BLR scales fall within the ensemble
distributions reported for nearby type-1 AGNs
\citep{Chen2023,Mandal2024DustTorus,Ayubinia2025}. Given the
uncertainties and intrinsic scatter of the scaling relations, the
differences between the two objects and between their $W1$ and $W2$
scales are not assigned independent physical significance.

This qualification is especially relevant for NLSy1 galaxies, whose high accretion rates can produce offsets from a single canonical $R_{\rm BLR}$--$L_{5100}$ relation. For Mrk~42, correcting the SDSS spectrum for host-galaxy contamination lowers the inferred AGN continuum luminosity and therefore the relation-based BLR scale. This highlights the importance of host-galaxy contamination when comparing BLR and hot-dust reverberation scales in low-luminosity AGNs.

For Mrk~493, the published direct H$\beta$ reverberation lag places
the $W1/W2$-emitting hot dust outside the main H$\beta$-emitting
region \citep{Du2015SEAMBH4}.
Both sources therefore place the hot-dust reverberation region
several BLR radii from the nucleus, following the canonical
disc--BLR--dust hierarchy of type-1 AGNs.
Our Mrk~493 lag agrees within the uncertainties with the
ICCF-based values reported by \citet{Mandal2024DustTorus}.
Differences among individual lag estimates likely reflect the
sparse WISE sampling and the broad correlation constraints it
produces.

\subsection{Colour behaviour and host-galaxy dilution}
\label{sec:disc_colour_host}

Both galaxies become bluer as their unresolved optical central
regions brighten, consistent with an increasing fractional
contribution from a blue variable AGN continuum. The measured
colour--magnitude slopes nevertheless describe the combined
nuclear and host-galaxy signal. A red, slowly varying stellar
component can strengthen the apparent bluer-when-brighter relation
as its fractional contribution decreases with increasing nuclear
brightness.

The colour-derived median spectral indices likewise characterise
the unresolved central region and should not be interpreted as
intrinsic AGN continuum indices. The flux--flux relations are less
sensitive to a constant additive component and provide a more
direct description of the correlated variable emission, although
slowly varying host and circumnuclear components can still affect
them.

The two sources differ in their observed $W1-W2$ behaviour.
Mrk~493 shows a statistically significant colour--magnitude
relation, indicating measurable evolution of its effective
short-wavelength infrared colour. No statistically significant
$W1-W2$ colour trend is detected in Mrk~42. Its positive
optical--infrared lag independently supports a delayed hot-dust
response. More
generally, two-band infrared colour evolution provides information
complementary to reverberation lags. Phase-resolved $W1-W2$
diagnostics can also constrain the temporal structure of the dust
response, as illustrated for the nuclear infrared transient
UGC~11487/WTP14adeqka
\citep{Izviekova2026UGC11487}.

Host dilution is particularly relevant in the WISE measurements,
whose aperture includes a larger fraction of the host galaxy than
the ZTF measurements. Non-variable stellar and extended
circumnuclear emission reduces the fractional amplitude of the
nuclear $W1/W2$ variations and can partly explain the weak
$W1-W2$ response in Mrk~42.

The approximately linear optical and infrared flux--flux relations
indicate broadly stable effective colours of the dominant
correlated variable components over the sampled timescales.
However, successive WISE epochs are separated by approximately
six months, so shorter-timescale spectral changes are suppressed
in the epoch-averaged $W1/W2$ relations.

\subsection{State-dependent optical response in Mrk~42}
\label{sec:disc_mrk42_flare}

The prominent optical flare of Mrk~42 provides a probe of
state-dependent continuum variability. The $g-r$ versus $g$
relation shows a steeper bluer-when-brighter slope during
the flare than in the quiet state, indicating a stronger colour
response as the optical continuum brightens.
The $r-i$ versus $r$ relation also becomes steeper
during the flare, although its uncertainty is larger because fewer
nightly pairs are available. This stronger state dependence suggests
an additional red contribution during the active interval.

The optical inter-band lag analysis supports this interpretation.
In the quiet state, the $g/r/i$ delays are consistent with zero
within their uncertainties. During the flare, the $i$-band response follows the $g$ and $r$
variations, with centroid delays of
$3.6^{+2.2}_{-2.7}$~d for $g$ versus $i$ and
$4.9^{+1.9}_{-3.4}$~d for $r$ versus $i$.
The steeper flare-state $r-i$ relation is therefore accompanied
by a measurable delayed contribution at the longest ZTF
wavelength.

Hot-dust emission extending into the $i$ band provides one
possible explanation. Torus contamination of the optical $I$ band
has been reported in Fairall~9
\citep{Mandal2026Fairall9}, demonstrating that dust-related
emission can contribute near the optical--infrared boundary. In
Mrk~42, a stronger and more coherent delayed response during the
flare could increase the fractional contribution of this red
component and steepen the observed $r-i$ relation. During the
quiet state, its weaker variability would be more strongly diluted
by host and other slowly varying emission.

The measurements therefore support a slowly responding red
component.
Hot-dust emission, diffuse low-ionization BLR emission, and
a broader disc-reprocessing response remain plausible within the
available cadence and filter coverage. Thermal-reverberation
calculations demonstrate that disc irradiation can produce
wavelength-dependent UV--optical continuum delays, with additional
diffuse BLR emission affecting particular bands
\citep{Kammoun2021,Kammoun2023}.

\subsection{Supplementary time-domain and multiwavelength constraints}
\label{sec:disc_supplementary}

The Swift monitoring of Mrk~493 provides complementary
information on its high-energy activity. The enhanced XRT interval
overlaps the broad active phase seen in the optical and UV light
curves, with a count-rate difference from the baseline significant
at $3.9\sigma$. Most of the enhancement is concentrated in the
soft band.

This soft-dominated behaviour is qualitatively consistent with
coupled variability in the inner accretion flow and corona.
\citet{Adegoke2019} measured a UV lead of approximately 5~ks
relative to the X-rays and interpreted it as a Comptonization delay
between disc seed photons and the hot corona. The soft-band
variability has also been modelled through changes in the
illuminating continuum, disc ionization, and relativistically
blurred reflection \citep{Bonson2018}. Broad-band
\textit{XMM--Newton}/\textit{NuSTAR} modelling revealed a
prominent soft excess and substantial reflection contribution
\citep{Tortosa2023}.

The present Swift behaviour is therefore compatible with
variability across the coupled disc--corona system. Both warm
Comptonization and variable reflection remain plausible within the
available cadence and spectral information. In this radial
interpretation, the optical, UV, and soft X-ray changes trace the
accretion-flow and coronal system, while the delayed $W1/W2$
response traces hot-dust reprocessing at larger radii.

The IAC80 observations provide an independent constraint on
hour-scale optical variability in Mrk~42 during three nights in
2025, several years after the prominent ZTF flare. The power-enhanced $F$-test detected no significant IDV during the
sampled quiescent intervals. These observations therefore provide
no evidence for hour-scale optical variability in the three
monitoring runs.

\section{Conclusions}
\label{sec:conclusions}

We presented an optical--infrared time-domain study of the isolated
narrow-line Seyfert~1 galaxies Mrk~42 and Mrk~493, based primarily
on ZTF and WISE/NEOWISE monitoring and supplemented by SDSS,
ATLAS, Swift, and IAC80 data. Because the survey
photometry samples unresolved central regions, the measured colours
and spectral indices are interpreted as combined nuclear,
host-galaxy, and circumnuclear quantities.

Both nuclei show positive optical--$W1/W2$ reverberation delays,
placing the responding hot dust at sub-parsec scales. The pyZDCF
and extended-driver tests support the positive-lag solutions, while
the Mrk~42 robustness analysis shows that its dust delays are
insensitive to the characteristic intra-flare optical structure.
Together, these results support delayed reprocessing of the variable
optical continuum by hot circumnuclear dust in both sources.

The inferred dust-response scales occupy the low-luminosity
extension of the published $R_{\rm dust}$--$L_{\rm bol}$
relations and place the hot-dust region several BLR radii from the
central source. For Mrk~493, the published H$\beta$ reverberation
lag independently places the $W1/W2$-emitting dust outside the main
H$\beta$-emitting region. For Mrk~42, correcting the SDSS spectrum
for host-galaxy contamination provides a more appropriate
luminosity-based BLR reference, illustrating the importance of host
subtraction when comparing BLR and dust-reverberation scales in
low-luminosity AGNs.

Both objects exhibit optical bluer-when-brighter behaviour, while
their $W1-W2$ responses differ. Mrk~493 shows significant
long-term evolution of its effective $W1-W2$ colour, whereas no
statistically significant colour trend is detected in Mrk~42.
The approximately linear flux--flux relations indicate broadly
stable effective colours of the dominant correlated variable
components. The different infrared colour behaviour also
illustrates the role of unresolved host and circumnuclear emission
in diluting the observed nuclear variability.

The major optical flare of Mrk~42 exhibits a characteristic
45--47~d intra-flare timescale. The non-uniform peak spacing,
red-noise tests, and Gaussian-process comparison favour structured
stochastic variability within a broader accretion-driven event,
without evidence for a coherent persistent oscillation. During the
flare, the delayed $i$-band response follows the expected
longer-wavelength ordering and indicates an additional slowly
responding red component; hot dust, diffuse low-ionization BLR
emission, and extended disc reprocessing remain plausible
contributors.

Overall, Mrk~42 and Mrk~493 follow a common
disc--BLR--hot-dust reverberation framework while displaying
source-specific optical flare and infrared colour behaviour.
The present analysis establishes a methodological framework for
extending optical--infrared reverberation studies to the broader
population of isolated AGNs in the 2MIG catalogue.

\section*{Acknowledgements}

We thank the anonymous referee for their constructive comments,
which helped improve the clarity of the manuscript.
This study was supported by the research programme for young
scientists of the National Academy of Sciences of Ukraine for
2025--2026 (Project ID 0125U002943). The authors acknowledge the
support of the Instituto de Astrofísica de Canarias (IAC).
The June 2025 visits of O.V. Kompaniiets and I.O. Izviekova to
the IAC, undertaken during the II EDUCADO Training School on
Computer Science, were co-funded by the European Union through the
MSCA Doctoral Network EDUCADO (GA 101119830) and the Widening
Participation project ExGal-Twin (GA 101158446). The March 2025
visit of I.O. Izviekova, I.B. Vavilova, and O.V. Kompaniiets was
supported by the EURIZON fellowship programme ``Remote Research
Grants for Ukrainian Researchers'' through Project ID 848,
``Evolutional properties of galaxies: classification by
morphological features, environmental influence, multiwavelength
properties of isolated galaxies with active nuclei''. The observations
were obtained with the IAC80 telescope, operated by the IAC at the
Teide Observatory, Tenerife, Spain. We thank the support astronomers
and night assistants for their assistance with the observations.

This work uses publicly available data from SDSS, ZTF,
\textit{NEOWISE}, and the NASA/IPAC Infrared Science Archive,
as well as Swift data products provided by the UK
Swift Science Data Centre at the University of Leicester.
The \textit{Neil Gehrels Swift Observatory} is a NASA mission with
participation from Italy and the UK. This research also made use of
NASA's Astrophysics Data System.

\section*{Data availability}

The NASA/IPAC Infrared Science Archive (IRSA) provides access to
\href{https://irsa.ipac.caltech.edu/docs/program_interface/ztf_lightcurve_api.html}{ZTF light curves}
and
\href{https://irsa.ipac.caltech.edu/Missions/wise.html}{WISE/NEOWISE photometry}.
ATLAS photometry is available from the
\href{https://fallingstar-data.com/forcedphot/}{ATLAS Forced Photometry Server}.
Target coordinates are listed in Table~\ref{tab:objects}.
The Mrk~42 spectrum is available through
\href{https://skyserver.sdss.org/}{SDSS SkyServer}.
Swift observations and XRT products for Mrk~493 are accessible through the
\href{https://www.swift.ac.uk/swift_portal/}{UK Swift archive} and the
\href{https://www.swift.ac.uk/user_objects/}{XRT product service}, respectively.

Reduced data from the IAC80/CAMELOT2 service-mode observations are
available from the corresponding author upon reasonable request.
The Mrk~42 comparison-star sequence (J2000 coordinates and APASS DR9
multiband photometry) and static finding chart are available in
\textit{AGN Reference Fields} \citep{Izviekova2026AGNfields} on Zenodo,
version~1.3
(\href{https://doi.org/10.5281/zenodo.23265726}{DOI 10.5281/zenodo.23265726}).
The chart and table are also available on the
\href{https://izviekova-blip.github.io/agn-reference-fields/mrk42.html}{Mrk~42 reference-field page}.

\bibliographystyle{mnras}
\begingroup
\linespread{0.92}\selectfont
\bibliography{library}
\endgroup

\appendix

\section{IAC80 observations and photometric calibration}
\label{app:iac80}

\begin{figure*}
\centering

\begin{minipage}{0.32\textwidth}
\centering
\includegraphics[width=\linewidth]
{\detokenize{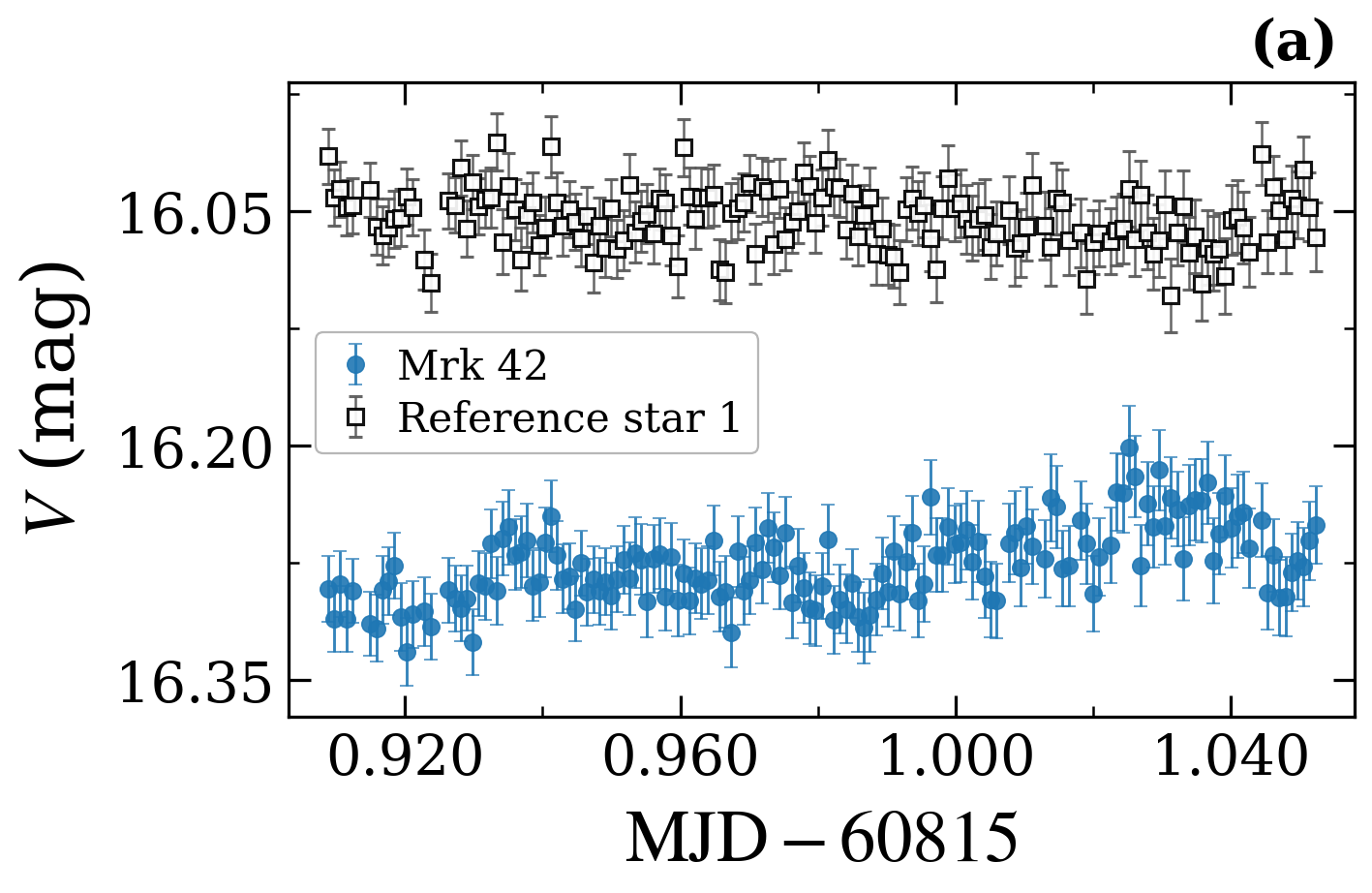}}
\end{minipage}
\hfill
\begin{minipage}{0.32\textwidth}
\centering
\includegraphics[width=\linewidth]
{\detokenize{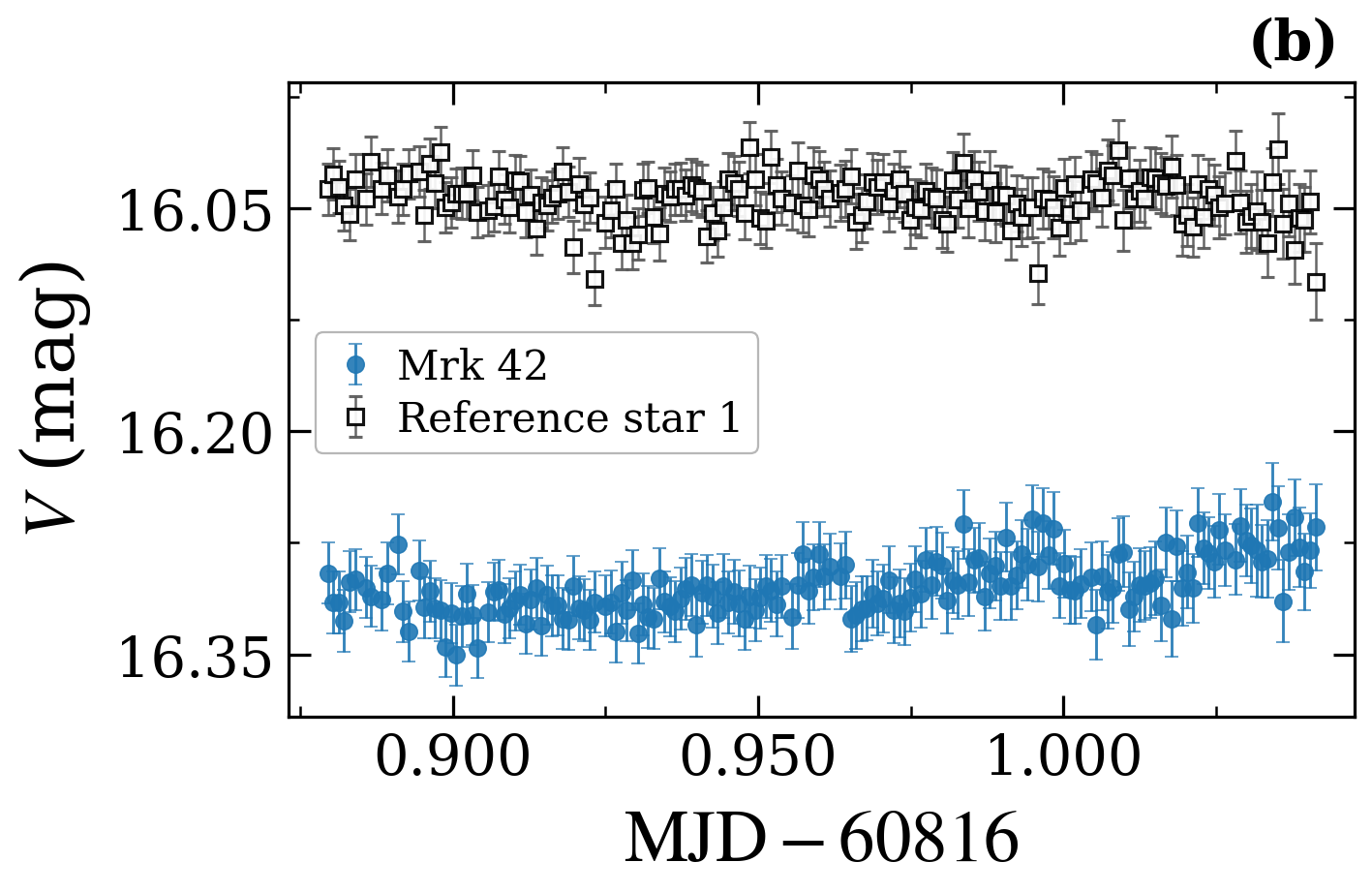}}
\end{minipage}
\hfill
\begin{minipage}{0.32\textwidth}
\centering
\includegraphics[width=\linewidth]
{\detokenize{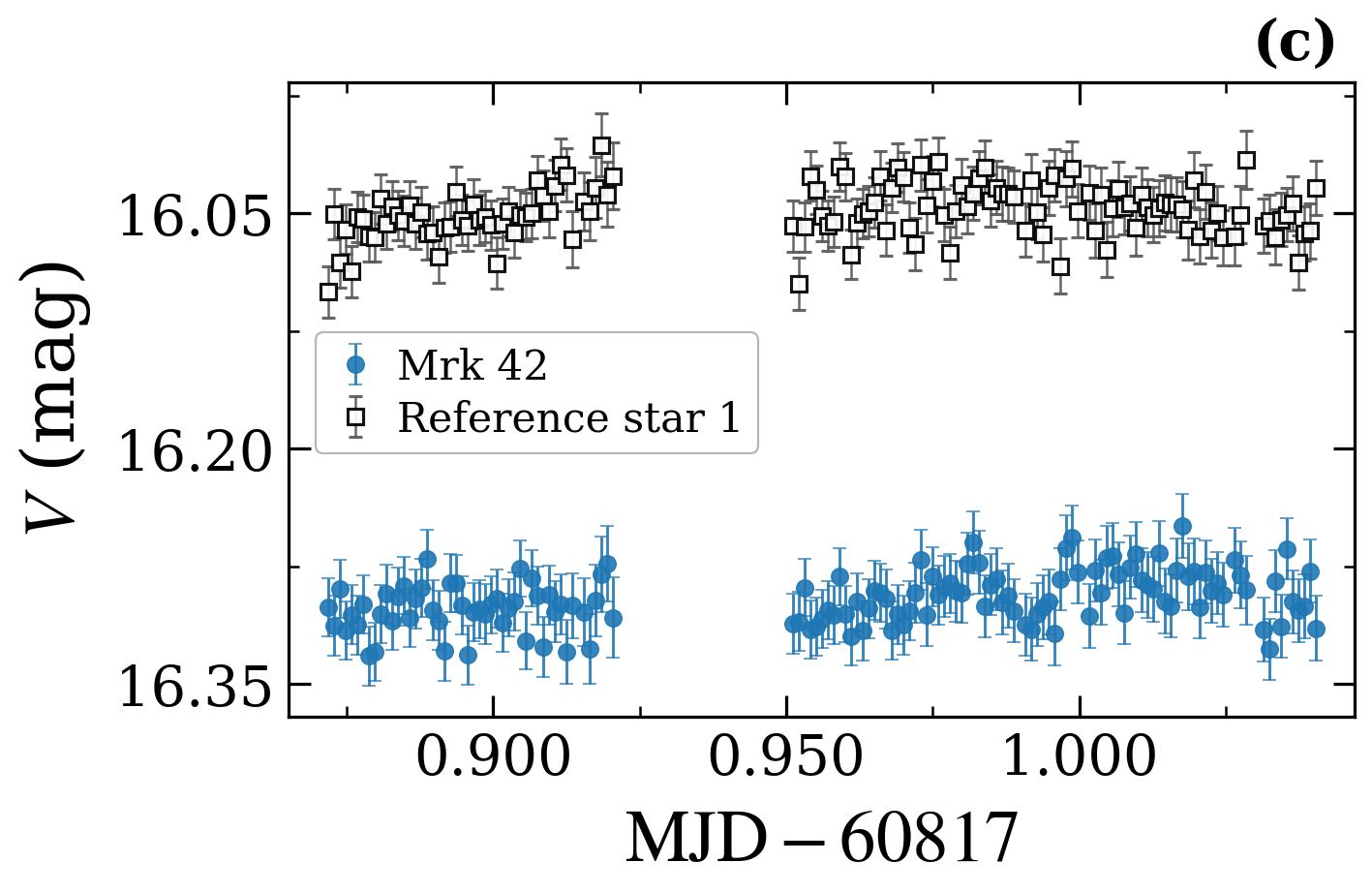}}
\end{minipage}

\caption{
IAC80 $V$-band IDV light curves of Mrk~42 measured using a
5-pixel ($1.61\arcsec$-radius) aperture. Panels (a), (b), and
(c) show the monitoring runs obtained on 2025 May 20/21, 21/22,
and 22/23, respectively. The Mrk~42 aperture light curve is shown
together with that of reference star~1 as a visual control. The
power-enhanced $F$-tests were performed using all four comparison
stars.
}
\label{fig:mrk42_idv}
\end{figure*}

Mrk~42 was monitored with IAC80/CAMELOT2 during three consecutive
nights in May 2025. The observing log and the resulting IDV light
curves are given in Table~\ref{tab:iac80_log} and
Fig.~\ref{fig:mrk42_idv}, respectively.
The comparison-star sequence used for the differential photometry,
including J2000 coordinates, APASS DR9 multiband catalogue
photometry, and a static finding chart, is publicly available in the
\textit{AGN Reference Fields} dataset
\citep{Izviekova2026AGNfields}.
The IAC80/CAMELOT2 $V$-band observations were analysed with
differential aperture photometry. Instrumental magnitudes were
measured for the central source and four local comparison stars
using the same circular aperture in every frame. The adopted
aperture radius was 5 pixels, corresponding to $1.61\arcsec$ for
the CAMELOT2 plate scale of
$0.322\arcsec\,{\rm pixel}^{-1}$. At the distance of Mrk~42, this
samples a projected central-region radius of approximately 0.8~kpc.

\refstepcounter{table}
\label{tab:iac80_log}

\medskip
\noindent
\textbf{Table~\thetable.}
Log of the IAC80/CAMELOT2 high-cadence monitoring observations of
Mrk~42 used in the intraday variability analysis.

\vspace{0.7ex}

\begin{center}
\footnotesize
\setlength{\tabcolsep}{3pt}
\renewcommand{\arraystretch}{1.08}

\begin{tabular}{@{}lcccccc@{}}
\hline
Night & Filter & $t_{\rm exp}$ & Start & End & Duration &
$N_{\rm frames}$ \\
      &        & (s) & (UTC) & (UTC) & (h) & \\
\hline
2025-05-20/21 & $V$ & 60 & 21:01:53 & 01:16:35 &
4.25 & $\sim254$ \\
2025-05-21/22 & $V$ & 60 & 21:06:32 & 01:04:32 &
3.97 & $\sim238$ \\
2025-05-22/23 & $V$ & 70 & 20:55:28 & 00:58:00 &
4.04 & $\sim207$ \\
\hline
\end{tabular}
\end{center}

\medskip

The local comparison sequence was selected using the
AAVSO Photometric All-Sky Survey Data Release~9 (APASS DR9) and
checked against Gaia Data Release~3 (Gaia DR3). Known or suspected
variable stars were excluded using the International Variable Star
Index (VSX) catalogue
\citep{Henden2016APASS,Vallenari2023GaiaDR3,Watson2006VSX}.
The star closest in brightness to the central source was used as the
reference star, and the remaining stars defined the pooled
comparison variance.
Intraday variability was assessed with the power-enhanced
$F$-test \citep{deDiego2014EnhancedF}. A night was classified as
variable when $F>F_{\rm crit}$ at a significance level of
$\alpha=0.01$.

\section{Lag-analysis diagnostics}
\label{app:lag_diagnostics}

\begin{figure*}
\centering

\begin{minipage}{0.48\textwidth}
\centering
\includegraphics[width=\linewidth]
{\detokenize{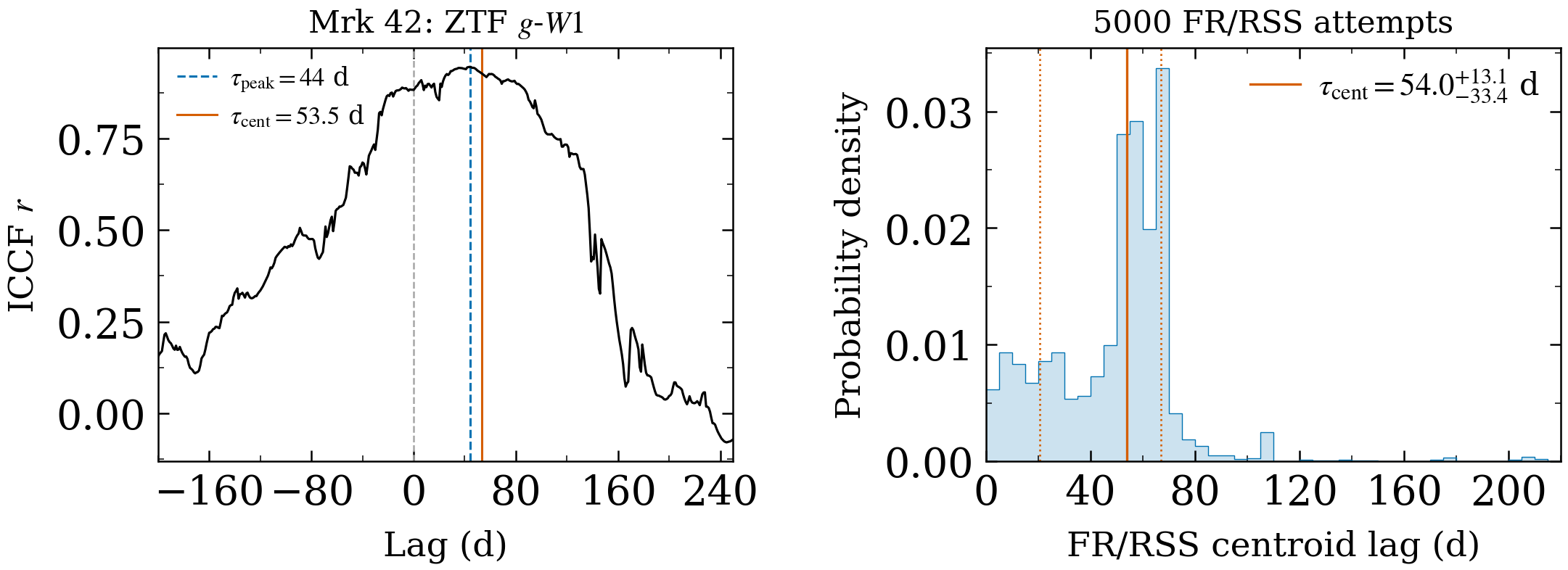}}\\[-0.5ex]
(a)
\end{minipage}
\hfill
\begin{minipage}{0.48\textwidth}
\centering
\includegraphics[width=\linewidth]
{\detokenize{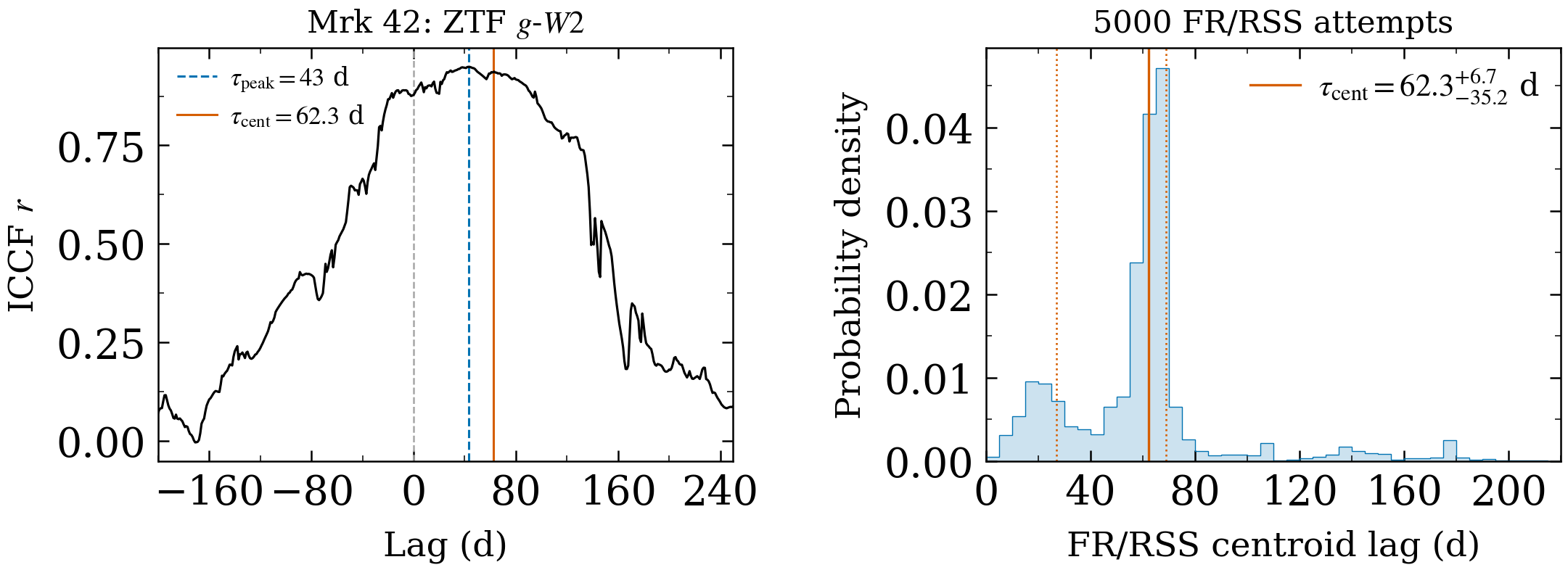}}\\[-0.5ex]
(b)
\end{minipage}

\caption{
ICCF and 5000-realisation FR/RSS analyses of the optical--infrared
delays in Mrk~42 using the accretion-disc-corrected infrared light
curves. Panel (a) shows the $g$--$W1$ analysis, and panel (b) shows
the $g$--$W2$ analysis. In each panel, the left-hand plot presents
the nominal ICCF, with the peak and centroid lags marked by vertical
lines, and the right-hand plot shows the FR/RSS centroid-lag
distribution. The solid vertical line marks the median centroid lag,
while the dotted lines mark the 16th and 84th percentiles.
}
\label{fig:app_mrk42_mirlags}
\label{fig:app_mrk42_gw1_lag}
\label{fig:app_mrk42_gw2_lag}
\end{figure*}

\begin{figure}
\centering
\includegraphics[width=\columnwidth]
{\detokenize{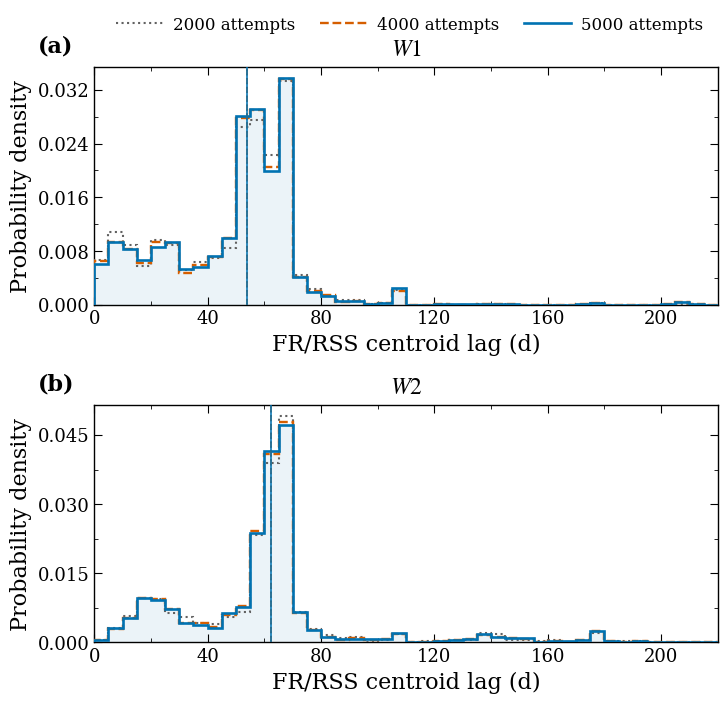}}

\caption{
Convergence of the FR/RSS centroid-lag distributions for Mrk~42.
Panels (a) and (b) show the $W1$ and $W2$ results, respectively,
after 2000, 4000, and 5000 FR/RSS realisations. The distributions
are normalised to unit area. Their close agreement shows that the
centroid estimates and 16th--84th percentile intervals are stable
with increasing numbers of realisations.
}
\label{fig:mrk42_frrss_convergence}
\end{figure}

\begin{figure}
\centering
\includegraphics[width=\columnwidth]
{\detokenize{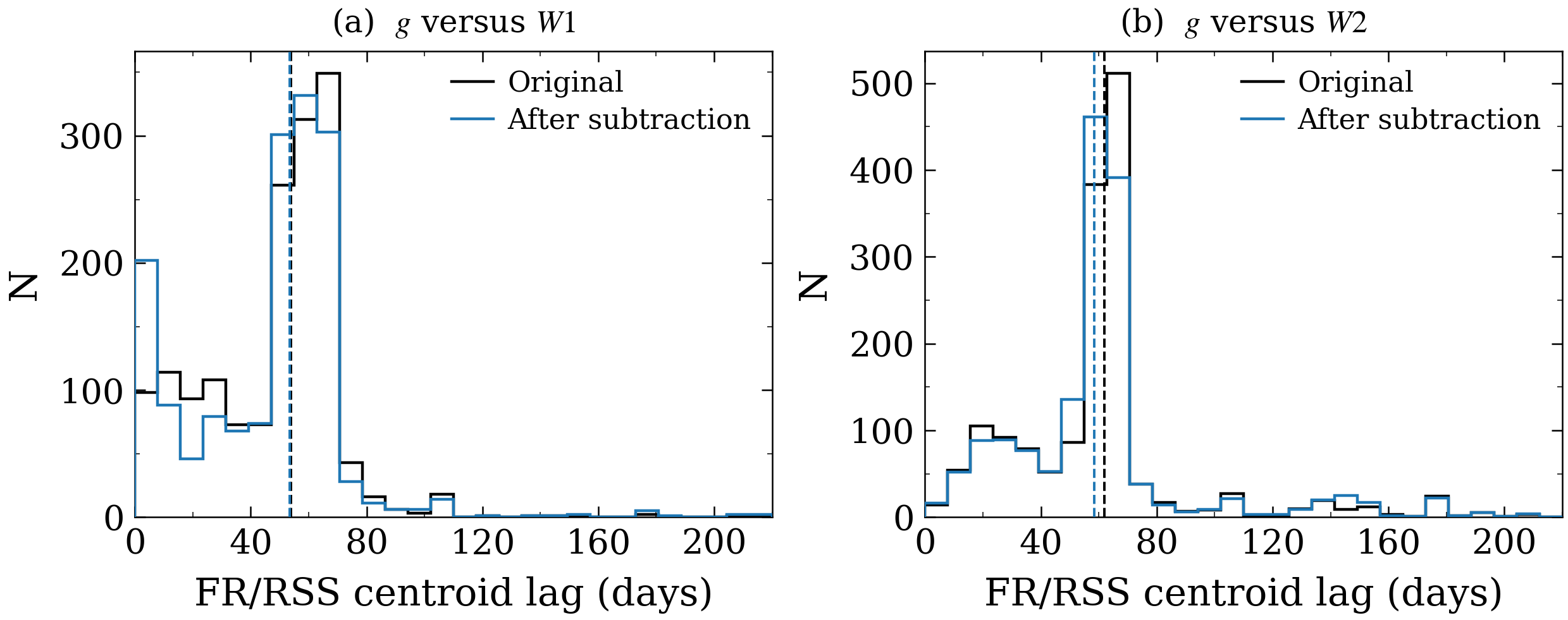}}
\caption{
FR/RSS centroid-lag distributions for Mrk~42 before
(black) and after (blue) subtraction of the fixed 47.12-d
residual component from the optical driver:
(a) $g$ versus $W1$ and (b) $g$ versus $W2$.
Dashed lines mark the distribution medians.
The paired centroid shifts are
$\Delta\tau_{\rm cent}=-1.0^{+1.0}_{-1.9}$~d
for $W1$ and
$-1.1^{+1.1}_{-2.2}$~d for $W2$, showing that
the inferred dust-reverberation delays are insensitive
to the residual intra-flare structure.
After subtraction, the centroid lags are
$54^{+12}_{-42}$~d for $W1$ and
$59^{+9}_{-30}$~d for $W2$.
}
\label{fig:mrk42_frrss_robustness}

\end{figure}

\begin{figure}
\centering

\begin{minipage}{\columnwidth}
\centering
\includegraphics[width=\linewidth]
{\detokenize{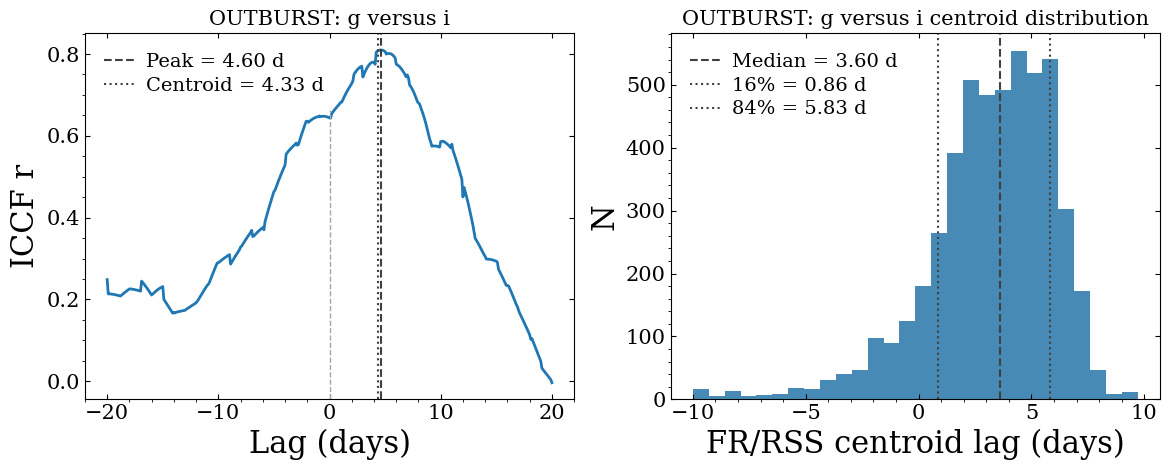}}\\[-0.5ex]
(a)
\end{minipage}

\vspace{1ex}

\begin{minipage}{\columnwidth}
\centering
\includegraphics[width=\linewidth]
{\detokenize{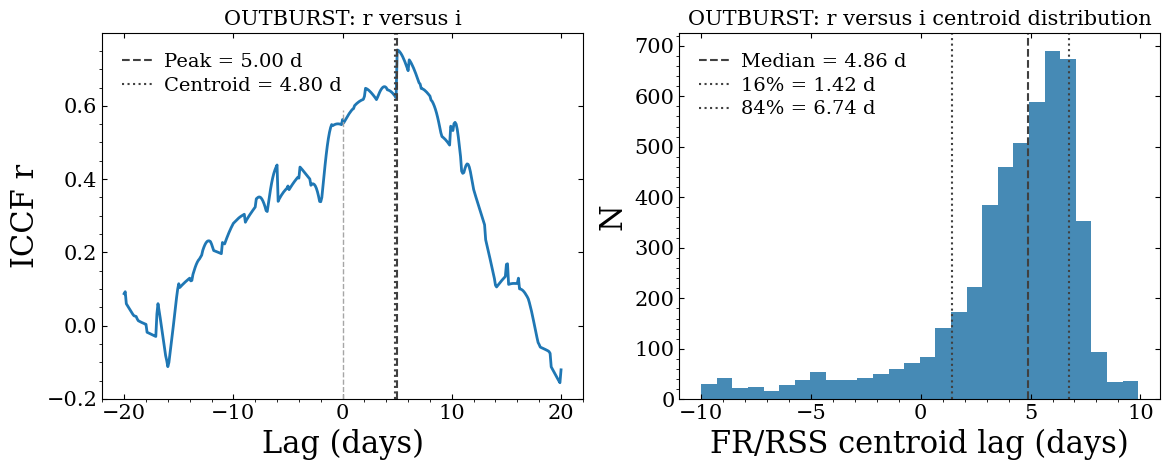}}\\[-0.5ex]
(b)
\end{minipage}

\caption{
Flare-state optical inter-band lag diagnostics for Mrk~42.
Panel (a) shows the $g$ versus $i$ analysis, and panel (b)
shows the $r$ versus $i$ analysis. Each filter pair was analysed with 5000 FR/RSS realisations.} The FR/RSS centroid
distributions indicate a positive response of the $i$ band relative
to the shorter-wavelength optical light curves.
\label{fig:app_mrk42_optlags}
\end{figure}

\begin{figure*}
\centering

\begin{minipage}{0.48\textwidth}
\centering
\includegraphics[width=\linewidth]
{\detokenize{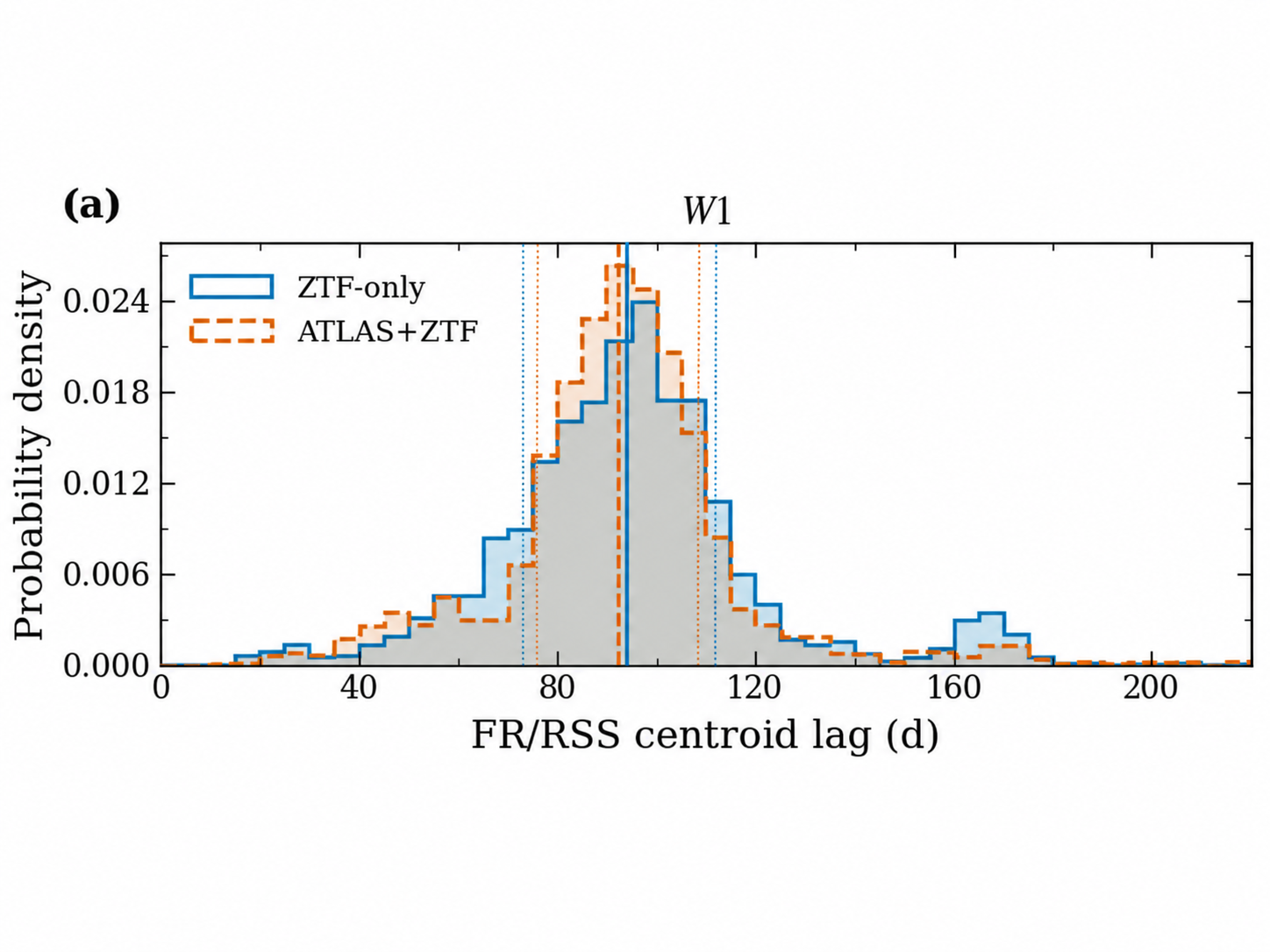}}
\end{minipage}
\hfill
\begin{minipage}{0.48\textwidth}
\centering
\includegraphics[width=\linewidth]
{\detokenize{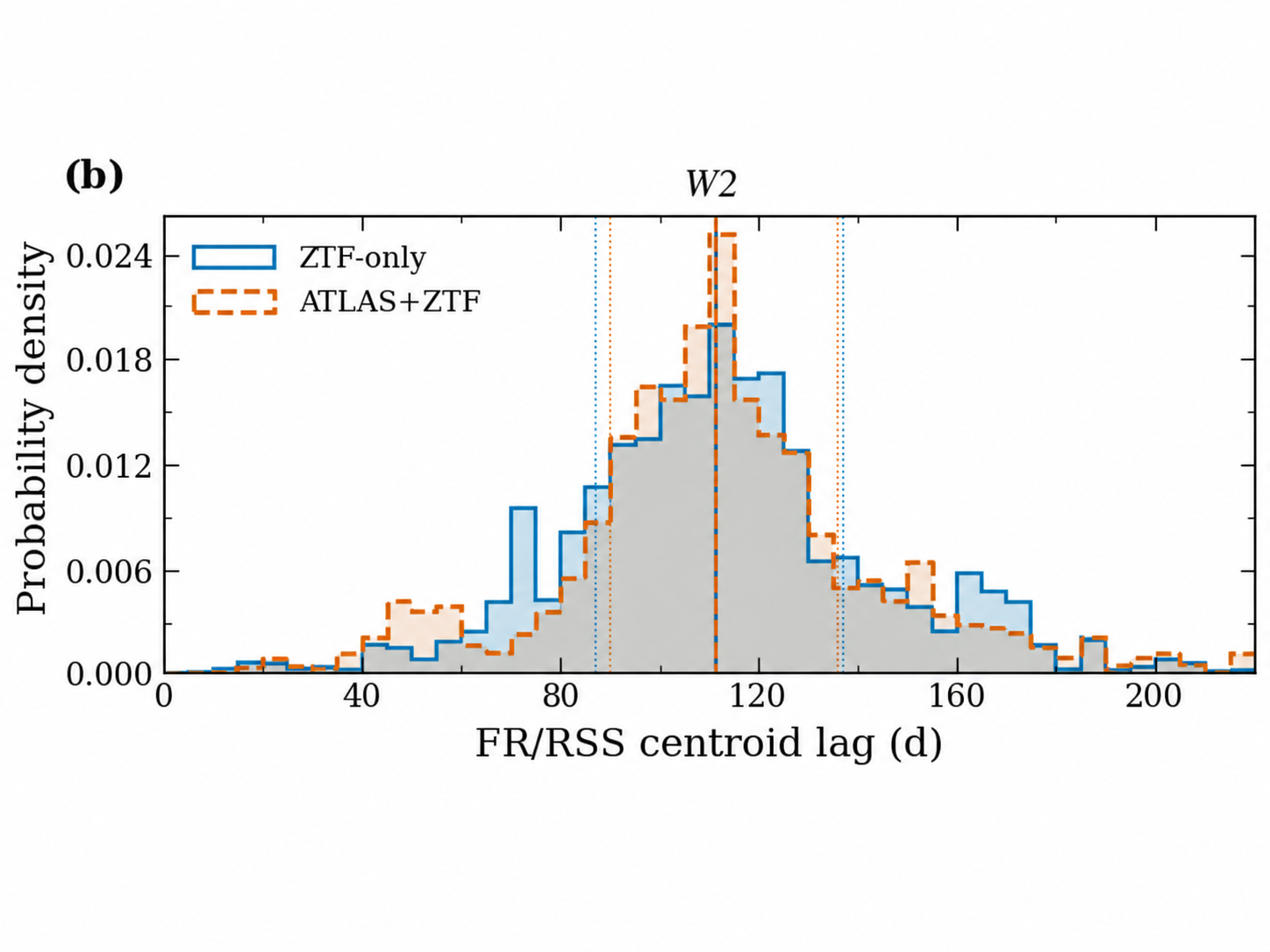}}
\end{minipage}

\caption{
FR/RSS centroid-lag distributions for Mrk~493 using the
fiducial ZTF-only (solid blue) and extended ATLAS+ZTF
(dashed orange) optical drivers: (a) $W1$ and (b) $W2$.
Solid vertical lines indicate median lags, and dotted lines
mark the 16th--84th percentile intervals.
The ZTF-only lags are $94^{+18}_{-21}$ and
$111^{+26}_{-24}$~d, while the extended-driver lags are
$93^{+16}_{-17}$ and $111^{+24}_{-21}$~d for $W1$
and $W2$, respectively.
}
\label{fig:app_mrk493_iccf}
\end{figure*}

The primary optical--infrared lag diagnostics for Mrk~42 are shown
in Fig.~\ref{fig:app_mrk42_mirlags}. The nominal ICCFs define the
peak and centroid positions, while the FR/RSS distributions provide
the fiducial centroid lags and their uncertainties.
The convergence of the fiducial Mrk~42 lag distributions with the
number of FR/RSS realisations is shown in
Fig.~\ref{fig:mrk42_frrss_convergence}.
Figure~\ref{fig:mrk42_frrss_robustness} compares the paired
FR/RSS centroid distributions obtained before and after subtraction
of the fixed 47.12-d residual component from the Mrk~42 optical
driver.
The flare-state optical inter-band diagnostics for Mrk~42 are
presented in Fig.~\ref{fig:app_mrk42_optlags}. These distributions
provide the lag measurements used in the analysis of the
state-dependent optical colour relations.
The stability of the Mrk~493 optical--infrared lag measurements
against extension of the optical baseline is illustrated in
Fig.~\ref{fig:app_mrk493_iccf}. The figure compares the FR/RSS
centroid distributions obtained with the fiducial ZTF $g$-band driver
and with the extended ATLAS+ZTF optical driver.

The pyZDCF calculations provide independent checks on the
optical--infrared correlation features recovered for both objects.
For Mrk~42, the formal pyZDCF maximum occurs at approximately
31~d in both bands and lies within the lower portions of the
corresponding FR/RSS ICCF intervals. The correlation functions
contain broad high-correlation plateaus extending across near-zero
and positive lags. Descriptive correlation-weighted centroids
calculated from the contiguous bins above $0.8r_{\rm max}$ are
approximately 17~d in both bands and are shifted toward shorter
lags by the near-zero part of the plateau. These descriptive centroids are not
adopted as independent reverberation measurements.

\begin{figure*}
\centering

\begin{minipage}{0.48\textwidth}
\centering
\includegraphics[width=\linewidth]
{\detokenize{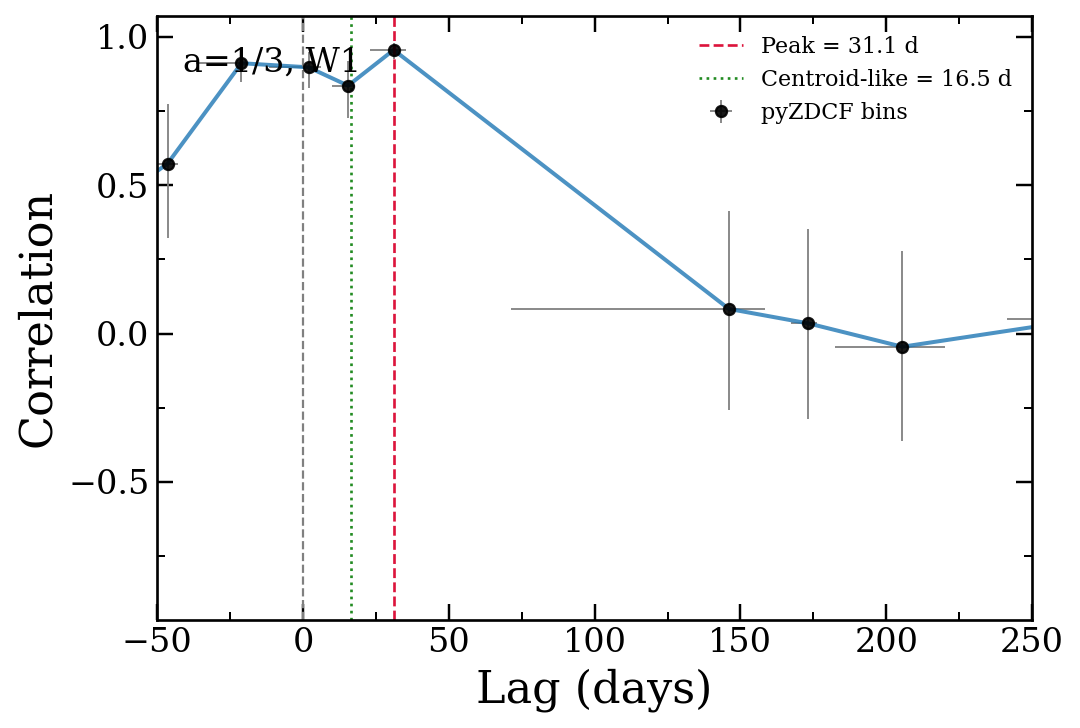}}\\[-0.5ex]
(a)
\end{minipage}
\hfill
\begin{minipage}{0.48\textwidth}
\centering
\includegraphics[width=\linewidth]
{\detokenize{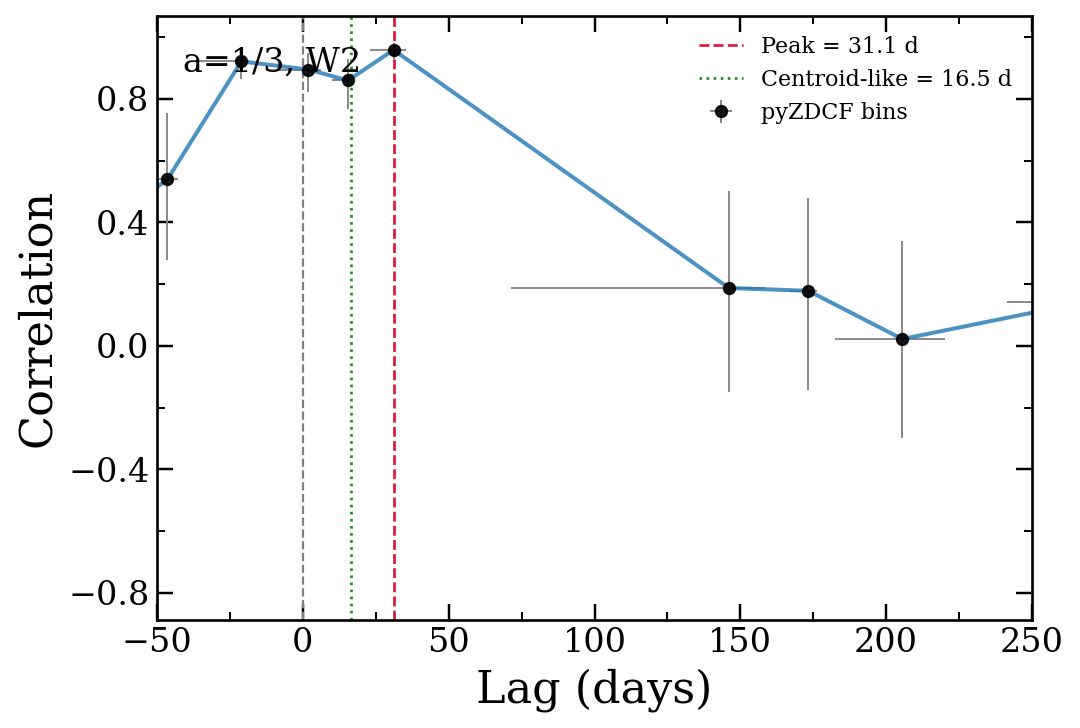}}\\[-0.5ex]
(b)
\end{minipage}

\vspace{1ex}

\begin{minipage}{0.48\textwidth}
\centering
\includegraphics[width=\linewidth]
{\detokenize{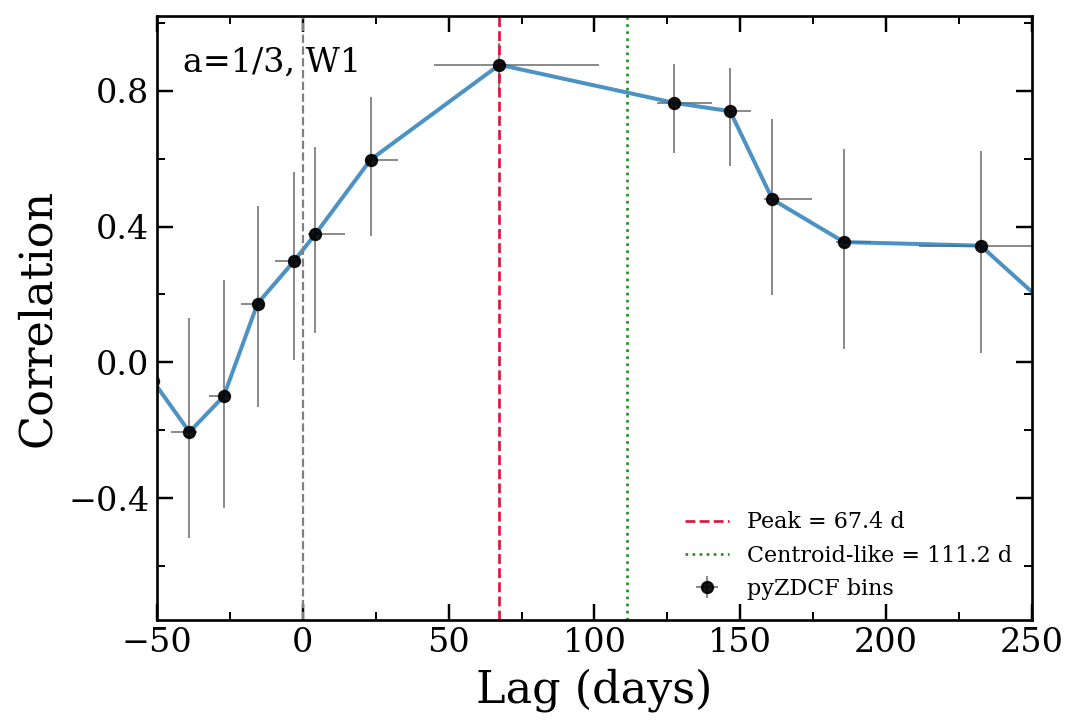}}\\[-0.5ex]
(c)
\end{minipage}
\hfill
\begin{minipage}{0.48\textwidth}
\centering
\includegraphics[width=\linewidth]
{\detokenize{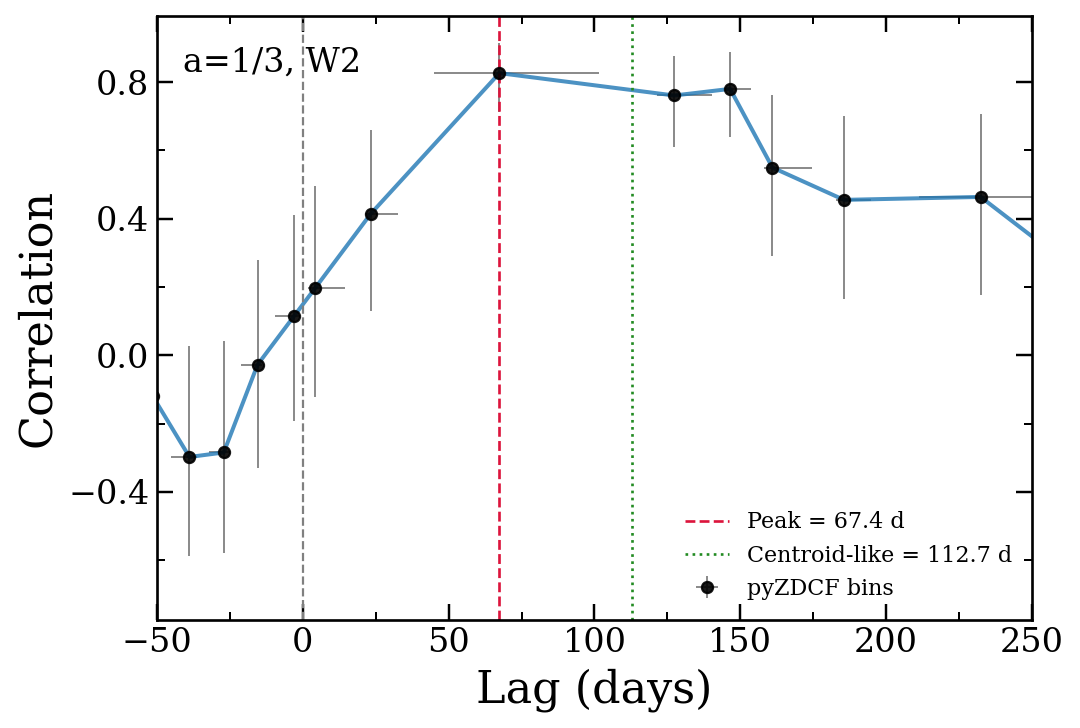}}\\[-0.5ex]
(d)
\end{minipage}

\caption{
pyZDCF consistency checks for the optical--infrared correlations
of Mrk~42 and Mrk~493. Panels (a) and (b) show the
$g$--$W1$ and $g$--$W2$ results for Mrk~42, respectively.
Panels (c) and (d) show the corresponding results for Mrk~493.
For Mrk~42, the red dashed lines mark the formal pyZDCF maxima
at approximately 31~d, and the green dotted lines mark the
descriptive correlation-weighted centroids of approximately 17~d. The broad
correlation features are used as independent checks on the
fiducial FR/RSS ICCF centroid measurements.
}
\label{fig:app_pyzdcf_all}
\label{fig:app_mrk42_pyzdcf}
\label{fig:app_mrk493_pyzdcf}
\end{figure*}

For Mrk~493, the pyZDCF calculations recover broad positive-lag
features compatible with the fiducial ICCF solutions. For both
objects, the FR/RSS ICCF centroid distributions remain the basis
of the reported reverberation lags. The corresponding pyZDCF diagnostics are shown in
Fig.~\ref{fig:app_pyzdcf_all}.

\section{Additional lag-analysis implementation details}
\label{app:lag_method_details}

For the Mrk~42 robustness test, the centroid-based ICCF and FR/RSS
analysis was repeated after subtracting a sinusoidal component with
a fixed period of 47.12~d from the optical driver over
MJD 58423.0--58692.0. The original and modified drivers were
compared using paired FR/RSS realisations with identical
random-subset selections and flux-randomisation deviates. The
effect of the subtraction was quantified from
\begin{equation}
\Delta\tau_{\rm cent}
=
\tau_{\rm cent,subtracted}
-
\tau_{\rm cent,original}.
\label{eq:paired_centroid_shift}
\end{equation}
Realisations without a finite centroid in the adopted positive-lag
interval were excluded from the corresponding distributions.

The pyZDCF calculations used a minimum of 11 points per bin and
300 Monte Carlo realisations \citep{Alexander1997}. Because of the
approximately six-month spacing of the NEOWISE epochs, the resulting
correlation functions provide broad lag constraints and are used as
independent consistency checks rather than as fiducial lag
measurements.

For the optical inter-band analysis of Mrk~42, the individual
cleaned ZTF $g$, $r$, and $i$ measurements were analysed
separately for the quiet and flare intervals.
The ICCF calculations used a lag range of $-20$ to $+20$~d
with a spacing of 0.1~d, a primary peak-search interval of
$-10$ to $+10$~d, a centroid threshold of $0.8\,r_{\rm max}$,
and at least ten overlapping pairs. The $g$ versus $r$,
$g$ versus $i$, and $r$ versus $i$ pairs were analysed with
the bidirectional ICCF procedure.
For each filter pair and activity state, 5000 FR/RSS
realisations were generated. The lag and its uncertainty were
derived from the 16th, 50th, and 84th percentiles of the
distribution of finite centroid estimates.
The pyZDCF calculations provided the corresponding independent
correlation check.

\section{Additional methodological details}
\label{app:additional_methods}

\subsection{Survey-photometry filtering}
\label{app:survey_processing}

For the ZTF photometry of Mrk~42, the major flare interval was
excluded from the automatic outlier rejection. The quiescent
baseline was cleaned using two-stage median-absolute-deviation
thresholds of $K_{\rm global}=2.0$ and $K_{\rm local}=3.0$ within
a 15-d moving window containing at least six neighbouring
measurements. All measurements within the adopted flare interval
were retained.

The ZTF colour slopes were additionally tested with stricter seeing
limits and after removing a fitted linear magnitude--FWHM
dependence, following \citet{Kompaniiets2026}. The resulting slopes
remained within the corresponding uncertainties, and
${\rm FWHM}\leq3.0\arcsec$ was retained for the fiducial analysis.

For NEOWISE, measurements were retrieved within $5\arcsec$ of the
adopted nuclear coordinates. We required finite magnitudes and
uncertainties, $\texttt{w1snr}>5$, $\texttt{w2snr}>3$,
$\texttt{qual\_frame}>0$, zero contamination or artefact flags in
the first two positions of $\texttt{cc\_flags}$, and
$\texttt{w1sat}=\texttt{w2sat}=0$.

Consecutive NEOWISE measurements separated by no more than 10~d
were assigned to the same observing epoch. Within each epoch,
robust cleaning was applied separately to $W1$ and $W2$ using local
and global median-absolute-deviation thresholds of 2.0, an
uncertainty threshold of 2.5 times the median epoch uncertainty,
and an interquartile-range threshold of 2.0 for epochs containing
more than four measurements. Clipping was suppressed when it would
leave fewer than three measurements, and the final sample retained
measurements accepted in both bands.

\subsection{Mrk~42 flare-substructure analysis}
\label{app:flare_method_details}

The internal structure of the major Mrk~42 flare was analysed in
the ZTF $g$- and $r$-band light curves over
MJD 58423.0--58692.0. Measurements within this interval were
excluded from the automatic median-absolute-deviation clipping
applied to the quiescent baseline and were retained for the
flare-substructure analysis.

The global flare morphology was characterised by comparing
Gaussian, asymmetric-Gaussian, and
exponential-rise/exponential-decay profiles using their AIC and BIC
values. The asymmetric Gaussian provided the preferred parametric
description of the broad flare envelope. For the residual timing
analysis, a weighted cubic smoothing spline was used as a flexible
non-parametric estimate of the low-frequency trend. The detrended
light curve was defined as
\begin{equation}
m_{\rm det}(t)=m(t)-T(t),
\label{eq:flare_detrending}
\end{equation}
where $T(t)$ is the fitted smooth trend.

Generalised Lomb--Scargle periodograms were calculated over
20--120~d, with the photometric uncertainties included as weights
\citep{Lomb1976,Scargle1982,Zechmeister2009,VanderPlas2018}.
The significance of the dominant periodogram feature was evaluated
with 3000 Timmer--Koenig red-noise surrogate light curves generated
with the observed sampling and measurement uncertainties
\citep{Timmer1995,Vaughan2005RedNoise}. The fiducial red-noise
slope was $\beta=1.05$, with $\beta=0.8$ and $\beta=1.3$ used to
assess sensitivity to the assumed noise spectrum.
For visualisation, a sinusoid with its period fixed to the dominant
$g$-band GLS timescale was fitted to the detrended residuals. The sinusoid is used to describe the residual structure
and to test its effect on the inferred reverberation lags.

The epochs of four broad $g$-band maxima were measured within
predefined local windows using the local light-curve morphology.
Timing uncertainties were estimated from the local sampling, and
the interval uncertainties were propagated from the corresponding
peak-time uncertainties. Their characteristic spacing was obtained
from
\begin{equation}
t_{\rm peak}(n)
=
t_{\rm ref}
+
\Delta t_{\rm fit}(n-1),
\label{eq:peak_spacing}
\end{equation}
where $n$ is the peak index. Because adjacent intervals share one
peak epoch, the three interval measurements are not statistically
independent. The fitted spacing is used only as a descriptive
timescale and is not interpreted as evidence for a coherent
periodic signal.

A time-domain Gaussian-process comparison was performed with
\texttt{celerite2} \citep{ForemanMackey2017Celerite}. The first
model used a stochastic exponential covariance term, while the
second included an additional damped simple-harmonic-oscillator
covariance term. The models were compared through their likelihoods,
AIC, and BIC \citep{Liddle2007}. These diagnostics quantify the
characteristic flare timescale and assess whether an additional
structured covariance component is required.

\subsection{Swift processing for Mrk~493}
\label{app:swift_processing}

The X-ray analysis used UKSSDC pipeline products in the
0.3--10~keV band \citep{Evans2007,Evans2009}. Individual
short-exposure snapshots were screened using criteria based on the
effective exposure time, fractional exposure, statistical
uncertainty, and approximate source-count level.
The cleaned XRT measurements were grouped into visit-like epochs
separated by $\Delta t>0.05$~d. Within each epoch, count rates were
calculated using inverse-variance weighting. Exposure-time-weighted
rates were also calculated as a consistency check. We examined the
total 0.3--10~keV count rate, the soft 0.3--1.5~keV rate, the hard
1.5--10~keV rate, and the hard-to-soft count-rate ratio,
$H/S=R_{\rm hard}/R_{\rm soft}$.

UVOT photometry was extracted from the sky images with the
\texttt{uvotsource} task \citep{Poole2008}
in HEASoft\footnote{\url{https://heasarc.gsfc.nasa.gov/docs/software/heasoft/}}
using calibration files from the Swift
CALDB\footnote{\url{https://swift.gsfc.nasa.gov/caldb/}}.
Source counts were measured within a
circular $3\arcsec$ aperture centred on the nucleus, and the
background was measured in nearby source-free regions.
We also compared photometry extracted using
$3\arcsec$ and $5\arcsec$ apertures, including the application
and omission of aperture corrections. The resulting differences
were within the photometric uncertainties for this spatially
extended source and did not affect the subsequent analysis.
The UVOT magnitudes were placed on the AB system using the
standard calibration and zero-points
\citep{Poole2008,Breeveld2011}.

For the multi-band comparison, the ZTF and UVOT photometry was
converted to flux density, while the XRT measurements were retained
as count rates. No temporal shifts were applied. Each series was expressed as a fractional deviation
from its own median,
\begin{equation}
\delta x=\frac{x-x_{\rm med}}{x_{\rm med}}.
\end{equation}

\label{lastpage}
\end{document}